\documentclass[preprints,review,accept,moreauthors]{Definitions/mdpi} 

\firstpage{1} 
\pubvolume{1}
\issuenum{1}
\articlenumber{0}
\pubyear{2025}
\copyrightyear{2025}
\datereceived{22 January 2026} 
\daterevised{20 February 2026} 
\dateaccepted{21 February 2026} 
\datepublished{ } 
\hreflink{https://doi.org/} 

\usepackage{bm, color, amsmath, amssymb}

\Title{Neutrino production mechanisms in strongly magnetized quark matter: Current status and open questions}

\TitleCitation{Neutrino production mechanisms in strongly magnetized quark matter}

\Author{Igor A. Shovkovy $^{1,2,\ddagger}$*\orcidA{} and Ritesh Ghosh$^{1,3,\ddagger}$*\orcidB{} }

\AuthorNames{ Igor A. Shovkovy and Ritesh Ghosh}

\AuthorCitation{Shovkovy, I. A.; Ghosh, R.}

\address{%
$^{1}$ \quad College of Integrative Sciences and Arts, Arizona State University, Mesa, Arizona 85212, USA\\
$^{2}$ \quad Department of Physics, Arizona State University, Tempe, Arizona 85287, USA
\\
$^{3}$ \quad Institute of Physics, Academia Sinica, Taipei 11529, Taiwan}

\corres{Correspondence: igor.shovkovy@asu.edu, riteshghosh283@gmail.com}

\secondnote{These authors contributed equally to this work.}

\abstract{We review the main neutrino emission mechanisms operating in dense quark matter under strong magnetic fields, with particular emphasis on conditions expected in the interiors of compact stars. We discuss the direct Urca and neutrino synchrotron processes in unpaired quark matter, incorporating the effects of Landau-level quantization. For the direct Urca process, the quantization of the electron energy spectrum plays a critical role, whereas quark quantization can often be neglected at sufficiently high baryon densities. The resulting field-dependent neutrino emissivity is anisotropic and exhibits an oscillatory behavior as a function of magnetic-field strength. We explore the implications of these effects for magnetar cooling and for possible anisotropic neutrino emission that could contribute to pulsar kicks. In addition, we review the $\nu\bar{\nu}$ synchrotron emission process, which, although subdominant, provides valuable insights into the interplay between magnetic fields and weak interactions in dense quark matter. Overall, our analysis highlights the nontrivial influence of strong magnetic fields on neutrino production in magnetized quark cores, with potential consequences for the thermal and dynamical evolution of compact stars.}


\date{20 Feb}

\begin{document}

\tableofcontents

\section{Introduction}
\label{sec:Introduction}

Compact stars provide a unique natural laboratory for studying matter under extreme conditions of density, temperature, and magnetic field strength. At densities exceeding a few times nuclear saturation, baryonic matter is expected to undergo a phase transition to a deconfined quark phase, possibly leading to the formation of quark cores in compact stars~\cite{Baym:2017whm,Annala:2023cwx}. The existence of such exotic matter in stellar interiors is of profound theoretical interest, as it can substantially alter the structural, thermal, and transport properties of compact stars.

Many compact stars, particularly magnetars, host extraordinarily strong magnetic fields, reaching up to $10^{15}~\mbox{G}$ at the surface and potentially much higher in their cores~\cite{Turolla:2015mwa,Kaspi:2017fwg,Lai_1991}. These intense fields can substantially modify the physical processes that govern stellar evolution, most notably neutrino emission, which dominates the cooling of young neutron stars over the first $10^5$--$10^6$ years~\cite{Yakovlev:2004iq}. Because neutrinos carry away the bulk of the thermal energy from the stellar interior, a detailed understanding of their production mechanisms is essential for interpreting observed cooling rates and constraining the properties of dense matter.

Magnetic fields influence neutrino emission through two main mechanisms: (i) by modifying reaction rates and (ii) by introducing anisotropies in the emission pattern. The first directly affects the overall cooling efficiency, while the second may contribute to large-scale dynamical effects such as pulsar kicks~\cite{Lai:1998sz,Arras:1998cv,Sagert:2007as}. These effects have been investigated extensively in nuclear matter~\cite{Baiko:1998jq,Riquelme:2005ac,Potekhin:2015qsa,Potekhin:2017ufy,Dehman:2022rpa,Tambe:2024usx,Kumamoto:2024jiq}. By contrast, systematic studies of neutrino emission in magnetized quark matter remained comparatively sparse~\cite{Xuewen:2005aa,Belyaev:2017nos,Ayala:2018kie,Ayala:2024wgb} until very recently~\cite{Ghosh:2025sjn,Ghosh:2025vkm}. This lack of comprehensive studies constituted a notable gap, particularly in light of the possibility that quark matter may be realized in some of the densest and most strongly magnetized astrophysical environments known.

In this report, we provide a comprehensive overview of recent advances in the quantitative description of neutrino emission from unpaired dense quark matter in strong magnetic fields. We focus in particular on detailed calculations of direct Urca and neutrino synchrotron emission rates, emphasizing the underlying theoretical framework developed in Refs.~\cite{Ghosh:2025sjn,Ghosh:2025vkm}.

\subsection{Weak processes producing neutrinos}
\label{sec:weak-processes}

\begin{figure}[tbh]
\centering
  \subfloat[\centering]{\includegraphics[height=0.175\textwidth]{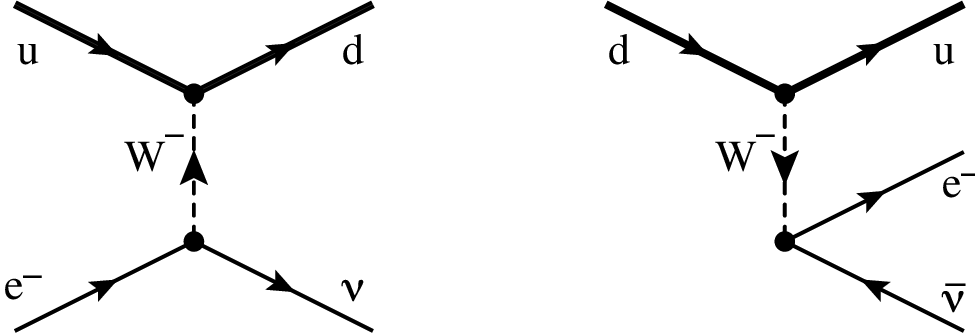}}
  \hspace{0.1\textwidth}
  \subfloat[\centering]{\includegraphics[height=0.175\textwidth]{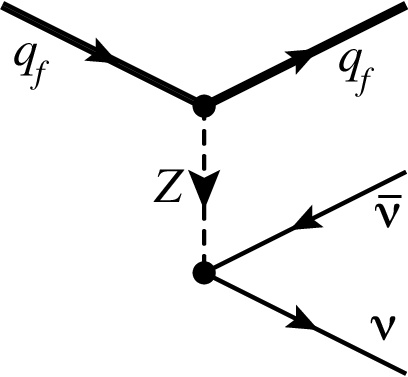}}
\caption{Feynman diagrams for (a) direct Urca (anti)neutrino emission and (b) neutrino-antineutrino synchrotron emission in magnetized quark matter.}
\label{fig.Feynman}
\end{figure}

In the absence of magnetic fields, the dominant neutrino emission mechanism in quark matter arises from the direct Urca processes,
\begin{equation}
d \rightarrow u + e^- + \bar{\nu}_e, \qquad 
u + e^- \rightarrow d + \nu_e,
\label{process:dUrca}
\end{equation}
which govern the cooling dynamics of unpaired quark matter~\cite{Iwamoto:1980eb,Iwamoto:1982zz}. The corresponding Feynman diagrams are shown in Fig.~\ref{fig.Feynman}(a). In Sec.~\ref{sec:dUrca-general}, we examine how strong magnetic fields modify the associated emission rates.
 
In addition to inducing  corrections to the direct Urca processes, magnetic fields also open new neutrino-emission channels that are otherwise forbidden by energy-momentum conservation. An important example is neutrino–antineutrino synchrotron emission,
\begin{equation}
q_f \rightarrow q_f + \nu_i + \bar{\nu}_i,
\label{process:synchrotron}
\end{equation}
where $q_f$ denotes a charged quark ($f = u, d$) and $\nu_i$ a neutrino of flavor $i = e, \mu, \tau$. This process, represented by the Feynman diagram in Fig.~\ref{fig.Feynman}(b), is associated with the cyclotron motion of charged fermions in a magnetic field. It has been studied extensively in the context of magnetized electron gases in white dwarfs and neutron stars~\cite{Yakovlev:1981AN,Kaminker:1992su,Kaminker:1993ey,Vidaurre:1995iv,Bezchastnov:1997ew}. In contrast, its analogue in quark matter has received far less attention.

The aim of this review is to present a coherent overview of neutrino emission mechanisms in magnetized quark matter, including both the modification of conventional direct Urca processes \cite{Ghosh:2025sjn} and field-induced neutrino-antineutrino synchrotron emission \cite{Ghosh:2025vkm}. We outline the theoretical framework based on the Kadanoff-Baym transport formalism, summarize the main theoretical results, discuss their implications for cooling and kick phenomena, and identify open questions relevant to astrophysical applications, particularly in the contexts of magnetars and hybrid stars. This discussion is intended to connect novel theoretical developments with observable phenomena, providing a comprehensive perspective on how strong magnetic fields influence the neutrino emissivity and thermal evolution of quark matter in compact stars.

\subsection{Setup and model assumptions}
\label{sec:model}

We consider unpaired two-flavor ($u,d$) quark matter in $\beta$-equilibrium and electrical neutrality at temperatures below the neutrino trapping threshold \cite{Prakash:1996xs}, such that $\mu_\nu=0$. The quark and electron chemical potentials satisfy the following relation: $\mu_d = \mu_u + \mu_e$. 

To obtain a rough estimate of the relative number densities of quarks and electrons, one may employ the thermodynamic relations for a non-interacting Fermi gas at $B=0$. For a fermionic species $f$ with degeneracy $g_f$, mass $m_f$, and chemical potential $\mu_f$, the number density at temperature $T$ is
\begin{equation}
n_f
=
\frac{g_f}{2\pi^2}
\int_0^\infty
\left(
\frac{1}{e^{(E_f-\mu_f)/T}+1}
-
\frac{1}{e^{(E_f+\mu_f)/T}+1}
\right)
p^2 \, dp ,
\end{equation}
where $E_f = \sqrt{p^2 + m_f^2}$. 
The first term corresponds to particles and the second to antiparticles. For cold dense matter relevant to compact stars ($T \ll \mu_f$), the antiparticle contribution is exponentially suppressed and the zero temperature limit provides a good approximation. The corresponding number density relation for a non-interacting Fermi gas  reads
\begin{equation}
n_f = \frac{g_f}{6\pi^2} \left(\mu_f^2-m_f^2\right)^{3/2} \approx  \frac{g_f\mu_f^{3}}{6\pi^2},
\label{eq:num-density}
\end{equation}
where, in the second expression, the fermion mass has been neglected. This approximation is sufficient for obtaining an order-of-magnitude estimate in the ultrarelativistic regime, $m_f \ll \mu_f$. For the quarks, the degeneracy factors are $g_u=g_d=2N_c=6$, accounting for two spin states and three color degrees of freedom, whereas for electrons $g_e=2$, reflecting spin degeneracy only. 
Imposing electric charge neutrality,
\begin{equation}
\frac{2}{3}n_u - \frac{1}{3}n_d - n_e = 0,
\end{equation}
and solving it together with beta equilibrium relation, $\mu_d = \mu_u + \mu_e$, one finds approximately $\mu_u\simeq 0.8\mu_d$ and $\mu_e\simeq 0.2\mu_d$. Since the number density in Eq.~(\ref{eq:num-density}) scales as $g_f \mu_f^3$, this implies $n_u \simeq 0.5\,n_d$ and $n_e \simeq 3\times10^{-3}\,n_d$ \cite{Shovkovy:2004me}. Although the electron density is small (with less than about one electron per $500$ quarks), electrons play a key role in neutrino emission.

In this review, we adopt representative model parameters appropriate for quark-matter cores in compact stars. Specifically, we consider quark chemical potentials $\mu_f$ of order $300~\mathrm{MeV}$ ($f=u,d$) and an electron chemical potential $\mu_e$ of order $50~\mathrm{MeV}$. We also assume that the temperature is sufficiently low to avoid neutrino trapping, i.e., $T \lesssim 5~\mathrm{MeV}$, ensuring that $T \ll \mu_e \ll \mu_f$. Generally, the magnetic field is assumed to be strong but realistic, $\sqrt{|eB|} \lesssim 25~\mathrm{MeV}$, corresponding to $B \lesssim 10^{17}~\mathrm{G}$.

Neutrino emission rates from the direct Urca and synchrotron processes shown in Fig.~\ref{fig.Feynman} are computed using the low-energy Fermi theory of weak interactions. The corresponding interaction Lagrangian densities are given by~\cite{Peskin:1995ev}
\begin{eqnarray}
\mathcal{L}^{\rm (W)} &=& \frac{G_F \cos \theta_C}{\sqrt{2}} \bar u \gamma^\mu(1-\gamma_5) d \, \bar{e} \gamma_\mu (1-\gamma_5)\nu_e,
\label{interaction-dUrca} 
\\
\mathcal{L}^{\rm (Z)} &=& \frac{G_F }{\sqrt{2}} \bar \psi_\nu \gamma^\mu(1-\gamma_5) \psi_\nu \, \sum_{f}\bar{\psi_f} \gamma_\mu (c_V^f-c_A^f\gamma_5)\psi_f,
 \label{interaction-synchrotron}
\end{eqnarray}
respectively, where $G_F\approx 1.166\times 10^{-11}~\mbox{MeV}^{-2}$ is the Fermi coupling constant and $\theta_C$ is the Cabibbo angle (note that $\cos^2\theta_C \approx 0.948$). In essence, these Lagrangian densities represent the point-like approximations of the weak interaction mediated by the charged $W^\pm$ and neutral $Z$ bosons. The vector $(c_V^f)$ and axial-vector $(c_A^f)$ couplings appearing in Eq.~(\ref{interaction-synchrotron}) are listed in Table~\ref{cVcA-table}.

\begin{table}[b]
\caption{Vector and axial-vector couplings in weak interactions for $\sin^2\theta_W = 0.231$.
\label{cVcA-table}}
	\begin{adjustwidth}{-0.2\extralength}{0cm}
		\begin{tabularx}{\fulllength}{lCCCCC}
			\toprule
			\textbf{Particle process}	& \bm{$c_V^f$}	& \bm{$c_A^f$}     & \bm{$(c_V^f)^2+(c_A^f)^2$ } & \bm{$(c_V^f)^2-(c_A^f)^2$}	& \bm{$c_V^f c_A^f$} \\
			\midrule
$u\to u+\nu_i+\bar{\nu}_i$ & $\frac{1}{2}-\frac{4}{3}\sin^2\theta_W$ &  $\frac{1}{2}$ & 0.287 &  -0.213 & 0.096 \\
$d\to d+\nu_i+\bar{\nu}_i$ & $-\frac{1}{2}+\frac{2}{3}\sin^2\theta_W$ &  $-\frac{1}{2}$ &  0.370 & -0.130 & 0.173\\
$e^{-}\to e^{-}+\nu_{\mu,\tau}+\bar{\nu}_{\mu,\tau}$*
& $-\frac{1}{2}+2\sin^2\theta_W$ &  $-\frac{1}{2}$ & 0.251 & -0.249 & 0.019 \\
$e^{-}\to e^{-}+\nu_{e}+\bar{\nu}_{e}$ $^\dagger $
& $\frac{1}{2}+2\sin^2\theta_W$ &  $\frac{1}{2}$ & 1.175 & 0.675 & 0.481 \\
			\bottomrule
		\end{tabularx}\\
	\noindent{\footnotesize{* Only neutral-current interaction contributes to synchrotron emission of muon and tau neutrinos.}}\\
    \noindent{\footnotesize{$^\dagger $ These effective couplings incorporate the combined contributions of neutral- and charged-current interactions in $\nu_e \bar{\nu}_e$ synchrotron emission \cite{Yakovlev:1981AN,Kaminker:1992su}.}}
	\end{adjustwidth}
\end{table}

\begin{figure}[h]
\centering
\includegraphics[height=0.175\textwidth]{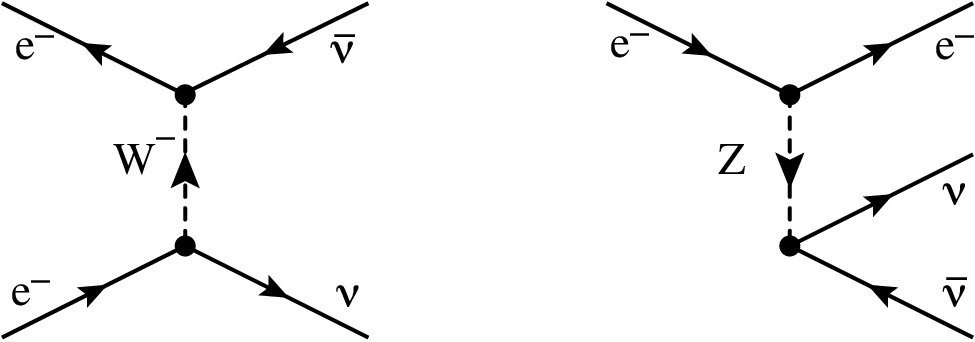}
\caption{The charged- and neutral-current diagrams contributing to synchrotron $\nu\bar{\nu}$ emission from electrons.}
\label{fig.electron.synchrotron}
\end{figure}

Let us mention in passing that synchrotron $\nu\bar{\nu}$ emission from electrons, unlike the synchrotron emission from quarks or heavy lepton flavors, is mediated by both charged- and neutral-current diagrams, as shown in Fig.~\ref{fig.electron.synchrotron}. As demonstrated in Refs.~\cite{Yakovlev:1981AN,Kaminker:1992su}, however, these two contributions lead to rate expressions with similar structures. The total result can therefore be expressed as a single contribution, provided that appropriate effective coupling constants $c_V^f$ and $c_A^f$ are used, see Table~\ref{cVcA-table}.

\subsection{Kadanoff-Baym transport equation for neutrinos}
\label{sec:Kadanoff-Baym-eq}

Instead of the traditional approach to neutrino emission, which relies on explicit amplitudes for the underlying tree-level electroweak processes represented by the interaction Lagrangians in Eqs.~(\ref{interaction-dUrca}) and (\ref{interaction-synchrotron}) and  shown in Fig.~\ref{fig.Feynman}, and which requires detailed knowledge of the cumbersome Landau-level wave functions in a background magnetic field~\cite{Yakovlev:1981AN,Kaminker:1992su}, we adopt an alternative framework, previously applied in different contexts~\cite{Sedrakian:1999jh,Schmitt:2005wg,Berdermann:2016mwt}. This framework is based on the Kadanoff-Baym formalism, which replaces explicit wave functions with Green’s functions and thus provides a more compact and computationally efficient treatment of neutrino emission processes.

Without loss of generality, we assume that the quark matter inside the stellar core is spatially uniform on macroscopic scales. Under this assumption, the nonequilibrium neutrino Green's functions $G^{\lessgtr}_\nu(t,P)$ depend only weakly on time and can be related to the (left-handed) neutrino distribution function as follows \cite{Sedrakian:1999jh,Schmitt:2005wg,Berdermann:2016mwt}:
\begin{eqnarray}
i G^{<}_\nu(t, P) &=& -\frac{\pi}{p}\frac{1-\gamma_5}{2} (\gamma^\lambda P_{\lambda}+\mu_\nu \gamma_0) \nonumber\\
&\times&\left\{f_\nu(t,\bm{p})\delta(p_{0}+\mu_\nu-p) 
-\left[1-f_{\bar \nu}(t,-\bm{p})\right]\delta(p_{0}+\mu_\nu+p)\right\}, 
\label{G-less} \\[1ex]
i G^{>}_\nu(t,P) &=& \frac{\pi}{p}\frac{1-\gamma_5}{2} (\gamma^\lambda P_{\lambda}+\mu_\nu \gamma_0)\nonumber\\
&\times&\left\{\left[1-f_\nu(t,\bm{p})\right]\delta(p_{0}+\mu_\nu-p)
-f_{\bar \nu}(t,-\bm{p})\delta(p_{0}+\mu_\nu+p)\right\}.
\label{G-greater}
\end{eqnarray}
Here $P =(p_{0},\bm{p})$ denotes the four-momentum of the neutrino, and $p =|\bm{p}|$ is the magnitude of its spatial momentum. Throughout this work, we consider conditions where neutrinos are not trapped in quark matter, so their chemical potential vanishes ($\mu_\nu =0$). To avoid potential confusion, we also note that, throughout this review, the subscript $\nu$ is used exclusively to denote a neutrino and never as a Lorentz index.

The nonequilibrium dynamics of neutrinos are governed by the Kadanoff-Baym transport equation,
\begin{equation}
i \partial_t \, \text{Tr}\!\left[\gamma^0 G_\nu^{<} (t,P)\right]
= -\,\text{Tr}\!\left[G_\nu^{>} (t,P)\Sigma^{<}_\nu(t,P)
-\Sigma^{>}_\nu(t,P)G_\nu^{<} (t,P)\right],
\label{KB-kinetic-eq}
\end{equation}
where $\Sigma^{\lessgtr}_\nu(t,P)$ denote the corresponding neutrino self-energies. The left-hand side encodes the time evolution of the neutrino distribution function included in $G_\nu^{<}(t,P)$, while the right-hand side represents an implicit collision term that encapsulates all information about electroweak interactions in Fig.~\ref{fig.Feynman} through the neutrino self-energies $\Sigma^{\lessgtr}_\nu(t,P)$.

Since the macroscopic evolution of dense matter occurs on timescales much longer than those characteristic of weak interactions, the explicit time dependence of the self-energies and Green's functions on the right-hand side of Eq.~(\ref{KB-kinetic-eq}) may be neglected. In certain astrophysical settings, such as core-collapse supernovae and neutron star mergers, the dynamical timescales may become comparable to those of weak processes~\cite{Alford:2017rxf,Arras:2018fxj,Most:2022yhe}. While this could appear to invalidate our approximation, this is not necessarily the case. If the average temperature and the quark and electron chemical potentials vary sufficiently slowly, it remains justified to neglect the time dependence on the right-hand side of Eq.~(\ref{KB-kinetic-eq}). This assumes, of course, that the other assumptions of our framework, i.e., low temperature and the absence of neutrino trapping, remain valid.

As we demonstrate in the following sections, this Kadanoff-Baym formalism provides a convenient starting point for calculating neutrino emissivities associated with weak processes, such as direct Urca reactions and neutrino-antineutrino synchrotron emission.

\section{Neutrino emission from direct Urca processes}
\label{sec:dUrca-general}

\begin{figure}[tbh]
\centering
\includegraphics[width=0.375\textwidth]{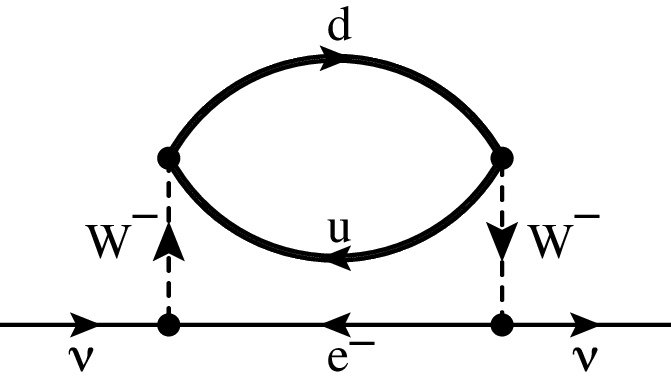}
\caption{Neutrino self-energy diagram used in the Kadanoff-Baym formalism to compute the direct Urca neutrino emission rate in dense quark matter.}
\label{fig.NuSelfEnergy}
\end{figure}

Using the interaction Lagrangian density in Eq.~(\ref{interaction-dUrca}), we obtain the following contribution to the neutrino self-energy,
\begin{equation}
\Sigma_\nu^{\lessgtr}(P_\nu) = i  \frac{G_F^2\cos^2\theta_C}{2} \int \frac{d^4 Q}{(2\pi)^4} \gamma^\delta (1-\gamma^5)\bar{S}_e^{\lessgtr}(P_\nu + Q)\gamma^\sigma(1-\gamma^5)\bar{\Pi}^{\gtrless }_{\delta\sigma}(Q) , 
\label{Sigma-gtr-1}
\end{equation}
which is represented by the Feynman diagram shown in Fig.~\ref{fig.NuSelfEnergy}. This self-energy effectively encapsulates the collision term in the Kadanoff-Baym equation (\ref{KB-kinetic-eq}).

Although the presence of a background magnetic field modifies the structure of the charged-particle Green's functions, the resulting expression for the neutrino self-energy in Eq.~(\ref{Sigma-gtr-1}) retains a form closely analogous to that in the zero-field case \cite{Sedrakian:1999jh,Schmitt:2005wg,Berdermann:2016mwt}. This can be understood as follows. In coordinate space, the electron Green's function and the $W$-boson self-energy can be written as
\begin{equation}
S_e^{\lessgtr}(u,u') = e^{i\Phi(u,u')}\bar S_e^{\lessgtr}(u-u'),
\qquad
\Pi^{\lessgtr}_{\delta\sigma}(u',u) = e^{-i\Phi(u,u')}\bar\Pi^{\lessgtr}_{\delta\sigma}(u'-u),
\label{S-and-Pi}
\end{equation}
where $\Phi(u,u')$ is the Schwinger phase and $u=(t,\bm r)$. Formally, the gauge-dependent Schwinger phases break translational invariance, thereby making a conventional momentum representation impossible. However, in the one-loop expression for the neutrino self-energy, which involves the product of the electron Green's function and the $W$-boson self-energy, the Schwinger phases cancel exactly. The remaining parts are translationally invariant and can therefore be Fourier transformed in the standard way
\begin{eqnarray}
\bar S_e^{\lessgtr}(P) &=& \int d^4u \, e^{iP^\lambda u_\lambda}\,\bar S_e^{\lessgtr}(u),
\label{S-Fourier}\\
\bar\Pi^{\gtrless}_{\delta\sigma}(Q) &=& \int d^4u \, e^{iQ^\lambda u_\lambda}\,\bar\Pi^{\gtrless}_{\delta\sigma}(u).
\label{Pi-Fourier}
\end{eqnarray}
Substituting these Fourier transforms into the neutrino self-energy expression then leads to Eq.~(\ref{Sigma-gtr-1}).

Thus, despite its conventional appearance similar to the zero-field case, Eq.~(\ref{Sigma-gtr-1}) fully incorporates the effects of the background magnetic field through the Fourier transforms of translationally invariant parts of Green's functions $\bar S_e^{\lessgtr}(P)$ and $\bar\Pi^{\gtrless}_{\delta\sigma}(Q)$. It should be stressed, however, that the transverse components of the four-momenta $P$ and $Q$ cannot be interpreted as physical momenta. For charged particles in a magnetic field, only the energy and the momentum component parallel to the field are conserved quantum numbers. The transverse dynamics, in contrast, is quantized into discrete Landau levels rather than described by continuous momenta.

The Fourier transform of the electron Green's functions can be written using a spectral representation as follows:
\begin{eqnarray}
i\bar S_e^{>}(P) &=& \left[1-n_F(p_0)\right] A_e(p_0+\mu_e,\bm p),
\label{S-gtr}\\
i\bar S_e^{<}(P) &=& -n_F(p_0)\,A_e(p_0+\mu_e,\bm p),
\label{S-less}
\end{eqnarray}
where $n_F(p_0)=\left[\exp(p_0/T)+1\right]^{-1}$ is the Fermi-Dirac distribution and $p_0$ is measured relative to the electron Fermi surface. The electron spectral function $A_e$ explicitly incorporates Landau-level quantization and is given by
\begin{eqnarray}
A_e(p_0+\mu_e,\bm p) &=& 2\pi\, e^{-p_\perp^2\ell^2}
\sum_{\lambda=\pm}\sum_{n=0}^{\infty}
\frac{(-1)^n}{E_{e,n}}
\delta(p_0+\mu_e-\lambda E_{e,n})\Big\{
\left[E_{e,n}\gamma^0-\lambda p_z\gamma^3+\lambda m_e\right] \nonumber\\
&\times&
\left[{\cal P}_+ L_n(2p_\perp^2\ell^2)
-{\cal P}_- L_{n-1}(2p_\perp^2\ell^2)\right]
+2\lambda(\bm p_\perp\!\cdot\!\bm\gamma_\perp)
L_{n-1}^1(2p_\perp^2\ell^2)
\Big\},
\label{spectral-density}
\end{eqnarray}
with $E_{e,n}=\sqrt{2n|eB|+p_z^2+m_e^2}$, $\ell=1/\sqrt{|eB|}$, ${\cal P}_\pm=(1\pm is_\perp\gamma^1\gamma^2)/2$, and $s_\perp = \mbox{sign}(eB)$.

The greater and lesser components of the $W$-boson self-energy are expressed in terms of the retarded self-energy as
\begin{eqnarray}
i\Pi_{\delta\sigma}^>(Q) &=& 2\left[1+n_B(q_0)\right]\mbox{Im\,}\Pi^R_{\delta\sigma}(Q),
\label{Pi-gtr}\\
i\Pi_{\delta\sigma}^<(Q) &=& 2n_B(q_0)\mbox{Im\,}\Pi^R_{\delta\sigma}(Q),
\label{Pi-less}
\end{eqnarray}
where $n_B(q_0)=\left[\exp(q_0/T)-1\right]^{-1}$ is the Bose-Einstein distribution function. The retarded self-energy $\Pi^R_{\delta\sigma}$ is generated by a one-loop quark diagram.

Although quarks are electrically charged, magnetic-field effects in quark propagators can be treated as weak and neglected to a good approximation in dense quark matter. For realistic magnetic fields $B\lesssim10^{17}\,\mathrm{G}$, the scale $\sqrt{|eB|}$ is much smaller than the quark chemical potentials $\mu_u,\mu_d\sim300\,\mathrm{MeV}$. Consequently, many quark Landau levels are occupied, and the level spacing near the Fermi surface, $\Delta\epsilon_n\sim|eB|/\mu_f$, is negligible compared with thermal smearing and interaction-induced broadening. This justifies neglecting the magnetic-field dependence of $\Pi^R_{\delta\sigma}$ at leading order.

In contrast, electrons have much smaller chemical potentials, $\mu_e\sim50\,\mathrm{MeV}$, making Landau-level quantization essential. For this reason, the electron propagator is treated exactly in the magnetic field through the spectral function in Eq.~(\ref{spectral-density}).

\subsection{Neutrino number production rate}
\label{sec:dUrca-derivation}

By substituting the neutrino Green’s function given in Eq.~\eqref{G-less}, the electron Green’s functions in Eqs.~\eqref{S-gtr} and \eqref{S-less}, expressed through the spectral function in Eq.~\eqref{spectral-density}, together with the $W$-boson self-energies from Eqs.~\eqref{Pi-gtr} and \eqref{Pi-less}, into the kinetic equation~\eqref{KB-kinetic-eq}, we derive the following expression for the neutrino number production rate:
\begin{eqnarray}
\frac{\partial f_\nu(t,\bm{p}_\nu)}{\partial t} &=& -\frac{G_F^2\cos^2\theta_C}{2 }
\sum_{\lambda=\pm}\sum_{n=0}^{\infty} (-1)^n
\int \frac{d^3\bm{p}_{e} e^{-p_{e,\perp}^2\ell^2} }{(2\pi)^3 p_\nu E_{e,n}}
n_F(E_{e,n}-\mu_e)  \nonumber\\
&\times & 
n_B(p_\nu+\mu_e-E_{e,n})  L_{n,\lambda}^{\delta\sigma}(\bm{p}_e,\bm{p}_\nu)  \mbox{Im} \left[ \Pi^R_{\delta\sigma}(Q) \right] ,
\label{rate-01}
\end{eqnarray}
where $Q\equiv(E_{e,n} -\mu_e-p_\nu,\bm{p}_e-\bm{p}_\nu)$. Note that the distribution function for antineutrinos 
satisfies a similar equation. The Landau-level dependent lepton tensor is defined by 
\begin{eqnarray}
L^{\delta\sigma}_{n,\lambda}(\bm{p}_e,\bm{p}_\nu)&=&\mbox{Tr}\Big[\Big\{
\left(E_{e,n}\gamma^{0} -\lambda  p_{e,z}\gamma^3+ \lambda m_e \right) 
\left[{\cal P}_{+}L_n\left(\xi_{p_e}\right)
-{\cal P}_{-}L_{n-1}\left(\xi_{p_e}\right)\right] \nonumber\\
&+&2\lambda  (\bm{p}_{e,\perp}\cdot\bm{\gamma}_\perp) L_{n-1}^1\left(\xi_{p_e}\right)
\Big\}\gamma^\sigma(1-\gamma^5)(\gamma_0 p_\nu-\bm{\gamma}\cdot \bm{p}_\nu)\gamma^\delta (1-\gamma^5)\Big].
\label{Lmunu}
\end{eqnarray}
where $\xi_{p_e} = 2 p_{e,\perp}^2\ell^2$. Using the explicit expression for the imaginary part of the $W$-boson self-energy, one finds that the contraction of the lepton and quark tensors, $L_{n,\lambda}^{\delta\sigma}(\bm{p}_e,\bm{p}_\nu)  \mbox{Im} \left[ \Pi^R_{\delta\sigma}(Q) \right]$, appearing on the right-hand side of Eq.~(\ref{rate-01}), is proportional to $\left(P\cdot P_\nu\right) \left(K\cdot Y_e\right)$. The latter closely resembles the zero-field particle amplitude \cite{Iwamoto:1980eb,Iwamoto:1982zz}. In addition to the four-momenta of the down and up quarks, $P$ and $K$, and the neutrino four-momentum $P_\nu$, this contraction depends on $Y_e$, which serves as an effective analog of the electron four-momentum. Unlike in the zero-field case, however, $Y_e$ encapsulates all information about the quantized Landau orbits. The explicit components of $Y_e$, are given by
\begin{eqnarray}
Y_{e,0} &=& \left(E_{e,n}-s_\perp \lambda p_{e,z}\right)L_{n}\left(\xi_{p_e}\right) 
-\left(E_{e,n}+s_\perp \lambda p_{e,z}\right)L_{n-1}\left(\xi_{p_e}\right) ,
\label{Y0-app}\\
Y_{e,z}&=&\left(\lambda p_{e,z}-s_\perp E_{e,n}\right)L_{n}\left(\xi_{p_e}\right) 
-\left(\lambda p_{e,z}+s_\perp  E_{e,n}\right)L_{n-1}\left(\xi_{p_e}\right) 
\label{Yz-app},\\
\bm{Y}_{e,\perp} &=& -4\lambda \bm{p}_{e,\perp} L_{n-1}^1\left(\xi_{p_e}\right).
\label{Yxy-app}
\end{eqnarray}
Evaluating the contraction $L_{n,\lambda}^{\delta\sigma}(\bm{p}_e,\bm{p}_\nu)  \mbox{Im} \left[ \Pi^R_{\delta\sigma}(Q) \right]$, where the effects of the magnetic field on the quark propagators are neglected, we obtain the following expression for the neutrino number production rate~\cite{Ghosh:2025sjn}:
\begin{align}
\frac{\partial f_\nu(t,\bm{p}_\nu)}{\partial t} &= \frac{N_c  G_F^2\cos^2\theta_C}{ 2  \pi^4}
\sum_{n=0}^{\infty} (-1)^n
\int \frac{pkdk d^3\bm{p}_{e} e^{-p_{e,\perp}^2\ell^2} }{v_Fq p_\nu E_{e,n}E_k E_p} \Theta(p_e) 
n_F(E_{e,n}-\mu_e) n_F( \mu_d-E_p)\nonumber\\
&\times  n_F(E_k-\mu_u)  \Bigg\{ \left[ E_p  p_{\nu,0}- \left(1+ \frac{k}{q} \cos\theta_{eu}\right) (\bm{q}\cdot \bm{p}_\nu) \right]\left( E_k Y_{e,0}  -\frac{k}{q}  \cos\theta_{eu} (\bm{q}\cdot \bm{Y}_{e}) \right) 
\nonumber\\
&+\frac{1-\cos^2\theta_{eu}}{2}\left( k^2 (\bm{p}_\nu\cdot \bm{Y}_{e}) -\frac{k^2}{q^2}(\bm{p}_\nu\cdot \bm{q})(\bm{Y}_{e}\cdot \bm{q})\right) 
\Bigg\} .
\label{rate-02}
\end{align}
In the derivation, we have used the energy conservation relation $p_\nu + \mu_e - E_{e,n} = E_k - \mu_u - E_p + \mu_d$. Also, we kept only the contribution from positive-energy electron states ($\lambda = 1$) and neglected the positron contribution ($\lambda = -1$), which is exponentially suppressed since the electron chemical potential is much larger than the temperature.

The explicit expression in Eq.~(\ref{rate-02}) reveals that the typical energies of the emitted neutrinos are of the order of the temperature, and that the quark states contributing most significantly to the direct Urca processes lie in a close vicinity of the Fermi surfaces. These kinematic constraints imply that the quark momenta $\bm{k}$ and $\bm{p}$, as well as the electron pseudo-momentum $\bm{p}_e$, are approximately parallel to one another. As in the zero-field case, such near-collinearity strongly suppresses the neutrino emission rate, particularly at low temperatures. However, as demonstrated by Iwamoto \cite{Iwamoto:1980eb,Iwamoto:1982zz}, this restriction is naturally relaxed once Fermi-liquid corrections to the quark dispersion relations are taken into account. These corrections reduce both the Fermi momenta and the Fermi velocities of the quarks, leading to the modified energy relations
\begin{eqnarray}
E_p &\simeq& \mu_d+v_F(p-p_F), \label{Ep-FL} \\
E_k &\simeq& \mu_u+v_F(k-k_F), \label{Ek-FL} 
\end{eqnarray}
where $p_F = v_F \mu_d$ and $k_F = v_F \mu_u$ are the Fermi momenta of the $d$ and $u$ quarks, respectively, and $v_F = 1 - \kappa$ is the Fermi velocity, with the strong-interaction effects captured in $\kappa = 2\alpha_s / (3\pi)$ \cite{Baym:1975va,Schafer:2004jp}. Note that the Fermi-liquid corrections for electrons are negligible (i.e., $p_{e,F} \simeq \mu_e$), since the QED fine-structure constant is much smaller than the QCD coupling $\alpha_s$.

After including the Fermi-liquid corrections, one finds that the three momenta are not collinear any more. Instead, the angle between the electron pseudo-momentum  $\bm{p}_e$ and the up quark momentum $\bm{k}$ is approximately determined by 
\begin{equation}
\cos\theta_{eu}  \simeq \frac{v_F^2 (\mu_d^2-\mu_u^2)-p_e^2}{2v_F\mu_u p_e} .
\label{cos-theta}
\end{equation}
This constraint has already been taken into account in deriving Eq.~(\ref{rate-02}) when performing the angular integrations associated with the direction of the $u$-quark momentum $\bm{k}$. To ensure that $|\cos\theta_{eu}| \leq 1$, the range of allowed values for the pseudo-momentum $p_e$ is restricted to $v_F\mu_e \leq p_e \leq v_F(\mu_d+\mu_u)$. This inequality is explicitly enforced in Eq.~(\ref{rate-02}) through the inclusion of unit-step functions inside $\Theta(p_e) $, which is defined as follows:
\begin{equation}
 \Theta(p_e)  \equiv \theta\left(p_e-v_F\mu_e\right)\theta\left[v_F(\mu_d+\mu_u)-p_e\right] .
\label{Theta-pe}
\end{equation}
While the effects of Fermi-liquid corrections are qualitatively similar to those encountered in the zero-field case~\cite{Iwamoto:1980eb,Iwamoto:1982zz}, important differences arise in the presence of a magnetic field. In particular, in Eq.~(\ref{cos-theta}) one can no longer assume that $p_e \equiv \sqrt{p_{e,\perp}^2 + p_{e,z}^2}$ represents the physical electron Fermi momentum, since $\bm{p}_{e,\perp}$ no longer corresponds to the physical transverse momentum and therefore does not uniquely determine the electron energy. As a result, the integration over $p_e$ in Eq.~(\ref{rate-02}) is not necessarily dominated by the region near $p_e \approx \mu_e$. 

Instead, the dominant contribution arises from the vicinity of the true electron Fermi surface, which in the presence of a magnetic field is organized into a discrete set of Landau levels. In particular, the patch of the Fermi surface associated with the $n$th Landau level is defined by longitudinal momenta satisfying $\sqrt{2n|eB|+p_{z,F}^2+m_e^2} = \mu_e$. The corresponding solutions are given by a discrete set of longitudinal momenta, $p_{z,F} = \pm \sqrt{\mu_e^2-m_e^2-2n|eB|}$, labeled by Landau-level indices $n\leq n_{\rm max}$, where $n_{\rm max}$ is given by the integer part of $(\mu_e^2-m_e^2)/(2|eB|)$. 

It is worth noting that the electron mass appears in the expression for the neutrino number production rate in Eq.~(\ref{rate-02}) only through the electron's energy $E_{e,n}$. However, since the rate is primarily determined by the states near the Fermi surface and the electron mass is much smaller than the chemical potential ($m_e\ll \mu_e$), its effect is negligible. Therefore, we can safely neglect it in our calculation below.

Further simplifications of the rate in Eq.~(\ref{rate-02}) can be made by replacing the quark momenta with their Fermi momenta throughout the integrand, except within the distribution functions, and by setting $\bm{q} = \bm{p}_e - \bm{p}_\nu \approx \bm{p}_e$. These approximations are well justified, since the dominant contribution to the rate originates from quarks within narrow energy bands ($\sim T$) around their Fermi surfaces, and the typical neutrino momenta are small compared to the electron pseudo-momentum ($T \ll v_F \mu_e \leq p_e $). Under these conditions, we obtain
\begin{align}
\frac{\partial f_\nu(t,\bm{p}_\nu)}{\partial t} &= \frac{N_c  G_F^2\cos^2\theta_C}{ 2  \pi^3}v_F \mu_u \mu_d
\sum_{n=0}^{\infty} (-1)^n
\int \frac{dp_{e,z} d(p_{e,\perp}^2) e^{-p_{e,\perp}^2\ell^2} }{p_e} 
 \Theta(p_e) \int dk n_F(E_k-\mu_u) \nonumber\\
& \times   
 n_F(p_{\nu}-E_k-E_{e,n}+\mu_d)n_F(E_{e,n}-\mu_e) \Bigg\{\left[ 1 - \left(1+ v_F \frac{\mu_u}{p_e} \cos\theta_{eu}\right) \frac{p_{e,z}p_{\nu,z} }{\mu_d p_{\nu}} \right] \nonumber\\
& \times   
\left( \frac{Y_{e,0}}{E_{e,n}}  -v_F \frac{(\bm{Y}_{e}\cdot \bm{p}_{e})}{p_e E_{e,n}} \cos\theta_{eu} \right)  
+ v_F^2 \frac{1-\cos^2\theta_{eu}}{2}\frac{\mu_u p_{\nu,z} }{\mu_d p_{\nu}}
\left( \frac{Y_{e,z}}{E_{e,n}} -\frac{p_{e,z} (\bm{Y}_{e}\cdot \bm{p}_{e})}{ p_e^2 E_{e,n}}\right) 
\Bigg\}, 
\label{rate-04}
\end{align}
where we have also performed the integration over the azimuthal angle $\phi_e$, specifying the direction of the electron pseudo-momentum $\bm{p}_{e,\perp}$ in the plane perpendicular to the magnetic field. Note that the scalar product $(\bm{Y}_{e} \cdot \bm{p}_{e})$ remaining in Eq.~(\ref{rate-04}) is independent of the angular coordinate $\phi_e$ and takes the following explicit form:
\begin{equation}
(\bm{Y}_{e}\cdot \bm{p}_{e}) = p_{e,z}^2 \left[L_{n}\left(\xi_{p_e}\right) -L_{n-1}\left(\xi_{p_e}\right)  \right]
-s_\perp E_{e,n} p_{e,z}\left[L_{n}\left(\xi_{p_e}\right) +L_{n-1}\left(\xi_{p_e}\right)  \right]  
-4 p_{e,\perp}^2 L_{n-1}^1\left(\xi_{p_e}\right).
\label{Ye-Pe}
\end{equation}

\subsection{Neutrino energy and momentum emission rates}
\label{sec:dUrca-rate}

The neutrino number production rate in Eq.~(\ref{rate-04}) can be used to determine the corresponding energy and net longitudinal momentum emission rates. These are defined as
\begin{eqnarray}
\dot{\cal E}_\nu &=& 2 \int \frac{d^3\bm{p}_{\nu}}{(2\pi)^3} p_{\nu,0} \frac{\partial f_\nu(t,\bm{p}_\nu)}{\partial t},
\label{dot-E-def}\\
\dot{\cal P}_{\nu,z} &=& 2 \int \frac{d^3\bm{p}_{\nu}}{(2\pi)^3} p_{\nu,z} \frac{\partial f_\nu(t,\bm{p}_\nu)}{\partial t}.
\label{dot-P-def}
\end{eqnarray}
The additional factor of $2$ accounts for the combined contributions from both direct and inverse Urca processes, $u + e^- \rightarrow d + \nu_e$ and $d \rightarrow u + e^- + \bar{\nu}_e$, one producing neutrinos and the other antineutrinos. 

A nonzero net longitudinal momentum emission rate, $\dot{\cal P}_{\nu,z}$, implies that neutrinos are emitted asymmetrically with respect to the direction of the magnetic field. Such an asymmetry is indeed expected in magnetized quark matter, where the combined effects of spin magnetization and parity violation induce a spatial imbalance in neutrino emission \cite{Sagert:2007as,Ayala:2018kie,Ayala:2024wgb}.

Substituting the neutrino number production rate from Eq.~(\ref{rate-04}) into Eqs.~(\ref{dot-E-def}) and (\ref{dot-P-def}), we obtain the following explicit expressions for the rates:
\begin{align}
\dot{\cal E}_\nu &= \frac{24 N_c G_F^2\cos^2\theta_C T^5}{\pi^5} v_F \mu_u \mu_d
\sum_{n=0}^{\infty} \frac{(-1)^n}{\ell^2}
\int_{0}^{1} dt \int_{v_F^2\mu_e^2\ell^2}^{v_F^2(\mu_d+\mu_u)^2\ell^2} ds \, e^{-s(1-t^2)}  
\Big\{L_{n}\left[2s(1-t^2)\right] \nonumber\\
& -L_{n-1}\left[2s(1-t^2)\right]  \Big\} g(\epsilon_n) 
\left[ 1 + \frac{ \sqrt{2n+st^2} }{2 \ell \mu_u} 
\left(1- \frac{v_F^2 \mu_e(\mu_d+\mu_u)\ell^2}{s}\right) \right]  , \label{rate-energy-a}
\end{align}
and 
\begin{align}
\dot{\cal P}_{\nu,z} &= s_\perp \frac{2N_c  G_F^2\cos^2\theta_C T^5}{\pi^5} v_F
\sum_{n=0}^{\infty} \frac{(-1)^n}{\ell^4}
\int_{0}^{1} dt \int_{v_F^2\mu_e^2\ell^2}^{v_F^2(\mu_d+\mu_u)^2\ell^2} ds \, e^{-s(1-t^2)}   \Big\{L_{n}\left[2s(1-t^2)\right] 
\nonumber\\
&+L_{n-1}\left[2s(1-t^2)\right]  \Big\}  g(\epsilon_n) 
\left[\frac{s(1-t^2)}{2}\left(1- \frac{v_F^2 (\mu_d+\mu_u)^2\ell^2}{s}\right)\left(1- \frac{v_F^2 \mu_e^2\ell^2}{s}\right) 
\right. \nonumber\\
& \left. +st^2\left(1 - \frac{v_F^2 \mu_e(\mu_d+\mu_u)\ell^2}{s} +\frac{2 \ell \mu_u}{\sqrt{2n+st^2} } \right) 
 \left(1+ \frac{v_F^2 \mu_e(\mu_d+\mu_u)\ell^2}{s} \right) 
\right] ,  
\label{rate-momentum-a}
\end{align}
respectively. In deriving these results, we performed the integration over the neutrino three-momentum $\bm{p}_{\nu}$, introduced the dimensionless integration variables $s = (p_{e,z}^2+p_{e,\perp}^2) \ell^2$ and 
$t = \sqrt{p_{e,z}^2/(p_{e,z}^2+p_{e,\perp}^2)}$, and adopted the following shorthand notations:
\begin{align}
g(\epsilon_n)&= \frac{1}{e^{\epsilon_n}+1} \left[ \mbox{Li}_{5}\left(e^{\epsilon_n} \right) -\frac{\epsilon_n}{4}\mbox{Li}_{4}\left(e^{\epsilon_n} \right)\right] ,\\
\epsilon_n&=\frac{\sqrt{2n+s t^2}-\mu_e\ell}{T\ell}.
 \end{align}
Note that the electron mass was neglected. This is justified since its value is much smaller than the electron chemical potential ($m_e\ll\mu_e$).

The direct Urca rates in Eqs.~(\ref{rate-energy-a}) and (\ref{rate-momentum-a}) constitute the main analytical results here. These expressions will be used in Sec.~\ref{sec:dUrca-numerical} to analyze the dependence of the energy and momentum emission rates on the magnetic field strength. It will be convenient to normalize our results to Iwamoto’s rate in the zero-field limit, which is reviewed in the next subsection.

\subsection{Zero magnetic field case}
\label{sec:zero-field-rate}

Using the same Kadanoff-Baym formalism, here we reproduce the well-known zero-field result originally derived by Iwamoto in the 1980s~\cite{Iwamoto:1980eb,Iwamoto:1982zz}. This serves as a useful benchmark for studying the neutrino emission rate in the presence of a nonzero magnetic field. We note in passing that several refinements of Iwamoto’s result, as well as extensions to color-superconducting quark matter (at $B=0$), can be found in Refs.~\cite{Schafer:2004jp,Jaikumar:2005hy,Schmitt:2005wg}.

Using the same neutrino Green’s function as in Eq.~(\ref{G-less}), together with the zero-field electron Green’s function and the corresponding $W$-boson self-energy, in the kinetic equation (\ref{KB-kinetic-eq}), we obtain the following expression for the neutrino number production rate \cite{Schmitt:2005wg}:
\begin{eqnarray}
\frac{\partial f_\nu(t,\bm{p}_\nu)}{\partial t} &=& \frac{N_c  G_F^2\cos^2\theta_C}{ 2  \pi^4}v_F \mu_u \mu_d
 \int \frac{dk d^3\bm{p}_{e} }{p_e}
n_F(E_{e}-\mu_e) 
\nonumber\\
&\times & n_F(p_{\nu}-E_k-E_{e}+\mu_d)n_F(E_k-\mu_u) \left( 1 -v_F \frac{p_e }{\mu_{e}}  \cos\theta^{(0)}_{eu}  \right) .
\label{rate-app1-B0}
\end{eqnarray}
This expression is the zero-field counterpart of Eq.~(\ref{rate-04}), which was derived for quark matter in the presence of a magnetic field.

Following the original approach of Refs.~\cite{Iwamoto:1980eb,Iwamoto:1982zz} to incorporate the Fermi-liquid corrections for quarks, the quark dispersion relations are replaced by those given in Eqs.~(\ref{Ep-FL}) and (\ref{Ek-FL}). The resulting kinematics constrain the angle between the electron and $u$-quark momenta to satisfy
\begin{equation}
\cos\theta^{(0)}_{eu}  \simeq \frac{v_F^2 (\mu_d^2-\mu_u^2)-p_e^2}{2v_F\mu_u p_e} 
\simeq v_F- (1-v_F^2) \frac{\mu_e}{2v_F\mu_u} ,
\label{costheta0}
\end{equation}
where, in the final approximation, $p_e$ has been replaced by the electron Fermi momentum, $p_{e,F} \simeq \mu_e$. Notably, if the Fermi-liquid corrections for quarks were neglected (i.e., $v_F \to 1$), $\cos\theta^{(0)}_{eu}$ would approach unity, indicating that the electron and up-quark momenta become collinear. Such a highly constrained phase space would result in a parametric suppression of the direct Urca process rates by a factor of $T / \mu_f$ \cite{Burrows:1980ec}.

The integrals in Eq.~(\ref{rate-app1-B0}) are dominated by contributions from particles near their respective Fermi surfaces, within an energy range of order $T$. Hence, we employ the standard approximation in which the quark and electron momenta are replaced by their Fermi momenta throughout the integrand, except within the distribution functions. Under this approximation, the neutrino number production rate becomes
\begin{eqnarray}
\frac{\partial f_\nu(t,\bm{p}_\nu)}{\partial t} &=& \frac{2N_c G_F^2\cos^2\theta_C}{ \pi^3}v_F \mu_u \mu_d \mu_e
 \int  dk dp_{e}  n_F(E_{e}-\mu_e) \nonumber\\
&\times &
n_F(p_{\nu}-E_k-E_{e}+\mu_d)n_F(E_k-\mu_u) \left( 1 -v_F \cos\theta^{(0)}_{eu} \right) .
\label{rate-app2-B0}
\end{eqnarray}
Noting that the antineutrino rate is the same, the total integrated energy rate reads
\begin{equation}
\dot{\cal E}^{\rm (Iwamoto)}_{\nu} =  2 \int \frac{d^3\bm{p}_{\nu}}{(2\pi)^3} p_{\nu,0} \frac{\partial f_\nu(t,\bm{p}_\nu)}{\partial t} 
\simeq \frac{457}{630} \alpha_s G_F^2\cos^2\theta_C \mu_u \mu_d \mu_e T^6 +O\left(\alpha_s^2,\frac{\mu_e}{\mu_u}\right),
  \label{dot-E-app-Iwamoto}
  \end{equation}
where we substituted $N_c=3$ and assumed that $\mu_e\ll \mu_u$. 

Some of the approximations used to derive the analytical rate in Eq.~\eqref{dot-E-app-Iwamoto} can be refined. In particular, one may retain the actual electron momentum $p_e$ throughout the integrand in Eq.~\eqref{rate-app1-B0}, rather than replacing it with the Fermi momentum $p_{e,F} \simeq \mu_e$. (A similar refinement for the $u$- and $d$-quark momenta is unnecessary, since their Fermi momenta $p_{u,F}$ and $p_{d,F}$ are much larger than $p_{e,F}$.) Numerical evaluation of the integral over $p_e$ with this improvement yields a slightly higher rate, which can be written in the modified form \cite{Ghosh:2025sjn}:
 \begin{equation}
 \dot{\cal E}_{\nu}^{(B=0)} \simeq C_T \frac{457\pi N_c}{2520}v_F (1-v_F^2)  G_F^2\cos^2\theta_C \mu_u \mu_d \mu_e T^6\left(1+\frac{\mu_e}{2\mu_u}\right),
 \label{dot-E-CT-app-B0}
\end{equation}
where $C_T$ is a function of order $1$, whose approximate dependence on the temperature is given by 
 \begin{equation}
C_T \approx 1+ c_1 \frac{T}{\mu_e}  +c_2 \frac{T^2}{\mu_e^2} , 
 \label{CT-app-B0}
\end{equation}
where $c_1\approx 15.70$ and $c_2\approx 6.287$.

\subsection{Numerical results for direct Urca rate}
\label{sec:dUrca-numerical}

Using the analytical expressions for the direct Urca rates derived in the previous two sections, we now present our numerical results for neutrino emission in magnetized quark matter. To reveal the effect of the background magnetic field, it is convenient to consider the dimensionless ratios $\dot{\mathcal{E}}_\nu / \dot{\mathcal{E}}_{\nu}^{(B=0)}$ and $\dot{\mathcal{P}}_{\nu,z} / \dot{\mathcal{E}}_{\nu}^{(B=0)}$, which characterize the energy and longitudinal momentum emission rates, respectively. Note that the latter ratio is normalized by the zero-field energy emission rate because the net momentum emission vanishes in the absence of a magnetic field, i.e., $\dot{\mathcal{P}}_{\nu,z}^{(B=0)}=0$.

\begin{figure}[t]
\centering
\includegraphics[width=0.9\textwidth]{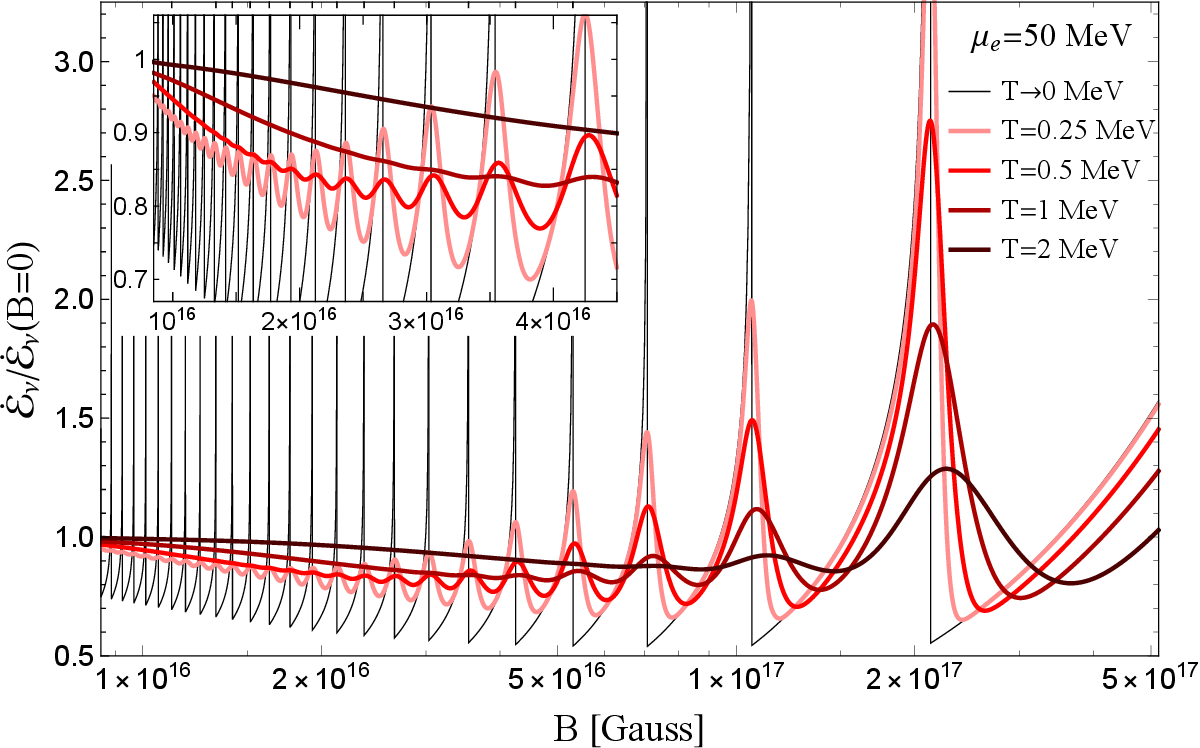}
\caption{The neutrino energy emission rates for four nonzero temperatures: $T = 0.25~\mbox{MeV}$, $0.5~\mbox{MeV}$, $1~\mbox{MeV}$, and $2~\mbox{MeV}$, together with the limiting case $T \to 0$. The inset shows a close-up view of the region with the magnetic field strength below $5\times 10^{16}~\mbox{G}$.\label{fig.energy.rates}}
\end{figure}

Our main numerical results for the dimensionless ratio $\dot{\mathcal{E}}_\nu / \dot{\mathcal{E}}_{\nu}^{(B=0)}$ as a function of the magnetic field are presented in Fig.~\ref{fig.energy.rates}. For illustrative purposes, we use the following values of the chemical potentials: $\mu_u = 300~\mathrm{MeV}$ and $\mu_e = 50~\mathrm{MeV}$. It should be noted, however, that the corresponding rates, when normalized to their zero-field values, are expected to depend only weakly on the specific values of the quark chemical potentials. To account for quark Fermi-liquid effects, we take the strong coupling constant to be $\alpha_s = 0.3$. 

In Fig.~\ref{fig.energy.rates}, we show numerical results for four representative nonzero temperatures: $T = 0.25~\mbox{MeV}$, $0.5~\mbox{MeV}$, $1~\mbox{MeV}$, and $2~\mbox{MeV}$, together with the limiting case $T \to 0$. As expected, the dimensionless ratios of the rates approach unity as $B \to 0$. To cover a broad range of magnetic fields, from approximately $|eB| \simeq 50~\mathrm{MeV}^2$ to $|eB| \simeq 5000~\mathrm{MeV}^2$, which is equivalent to the field strengths in a range from about $B\simeq 8.5\times 10^{15}~\mbox{G}$ to $B\simeq 8.5\times 10^{17}~\mbox{G}$, we use a logarithmic scale on the horizontal axis. Note that the squared-energy units of $|eB|$ can be converted to gauss using the relation $B = 1.69 \times 10^{14}~\mathrm{G}\, |eB|/\mathrm{MeV}^2$.

As see from Fig.~\ref{fig.energy.rates}, the rates exhibit a characteristic oscillatory dependence on the magnetic field, analogous to the de Haas-van Alphen effect in metals. As in the case of the magnetic susceptibility in metals, these oscillations arise from Landau-level quantization of electron states near the Fermi surface. Consequently, peaks in the energy emission rate occur when the Fermi energy coincides with the thresholds of individual Landau levels, i.e., when $|eB|/\mu_e^2 = 1/(2n)$ for positive integers $n$. These peaks become increasingly pronounced as the temperature decreases and formally diverge in the limit $T \to 0$. We note, however, that the energy emission rates in Fig.~\ref{fig.energy.rates} are normalized to their zero-field values given by Eq.~(\ref{dot-E-CT-app-B0}). Therefore, the absolute rates themselves vanish rapidly in the low-temperature limit, scaling as $\propto T^6$ when $T \to 0$.

Overall, the energy emission rate decreases with increasing magnetic field strength. However, even for fields as strong as $B \simeq 10^{17}~\mbox{G}$, the suppression is limited to only about $20\%$, suggesting that significant observational consequences are unlikely. An interesting exception arises in the lowest Landau level (LLL) regime, where the emission rate can be significantly enhanced. Realization of this regime requires extremely strong magnetic fields such that only the lowest Landau level is occupied, i.e., $eB \gtrsim \mu_e^2$, with $\mu_e$ denoting the electron chemical potential. For the representative value $\mu_e \simeq 50~\mbox{MeV}$ adopted here, this condition translates into $B \gtrsim 4.2 \times 10^{17}~\mbox{G}$. Although such field strengths are not excluded in principle, they lie near the upper end of theoretically plausible magnetic fields in stellar interiors, even when possible amplification mechanisms are taken into account \cite{Duncan:1992hi,Broderick:2001qw}.

\begin{figure}[t]
\centering
\includegraphics[width=0.9\textwidth]{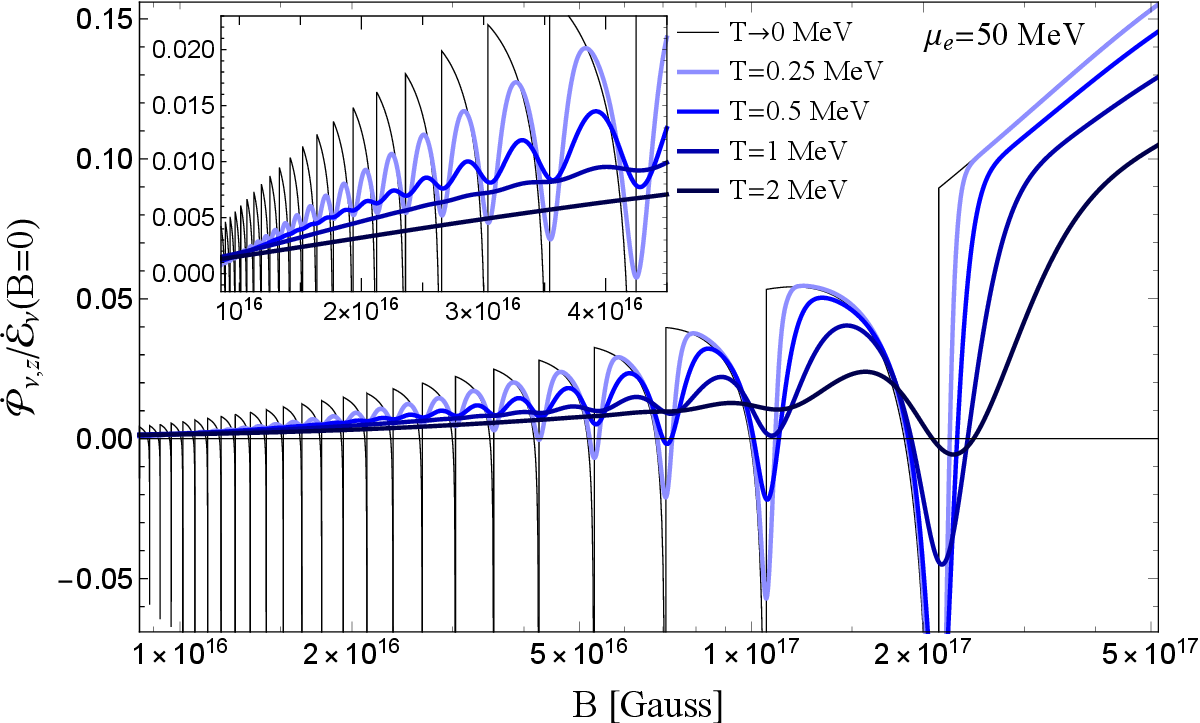}
\caption{The neutrino net momentum emission rates for four nonzero temperatures: $T = 0.25~\mbox{MeV}$, $0.5~\mbox{MeV}$, $1~\mbox{MeV}$, and $2~\mbox{MeV}$, together with the limiting case $T \to 0$. The inset shows a close-up view of the region with the magnetic field strength below $5\times 10^{16}~\mbox{G}$.\label{fig.momentum.rates}}
\end{figure}

Numerical results for the momentum emission rates are presented in Fig.~\ref{fig.momentum.rates} for the same four nonzero temperatures, $T = 0.25~\mbox{MeV}$, $0.5~\mbox{MeV}$, $1~\mbox{MeV}$, and $2~\mbox{MeV}$, together with the limiting case $T \to 0$. As before, we show the dimensionless ratio $\dot{\mathcal{P}}_{\nu,z} / \dot{\mathcal{E}}_{\nu}^{(B=0)}$ rather than the net longitudinal momentum emission itself. Similar to the energy emission rate, the net momentum exhibits an oscillatory dependence on the magnetic field, with the oscillation peaks becoming increasingly pronounced as the temperature decreases.

We note that, while the neutrino energy emission rate is strictly positive, the longitudinal momentum emission rate does not have a definite sign. In other words, the direction of the emitted net momentum is aligned with the magnetic field for certain values of $|eB|/\mu_e^2$ and opposite to it for others. As discussed in detail in Ref.~\cite{Ghosh:2025sjn}, this behavior arises from the angular dependence of the underlying weak processes. In particular, electrons near the equatorial region of the discretized Fermi surface ($p_{e,z} \simeq 0$) predominantly emit neutrinos in the direction opposite to those emitted by electrons near the polar regions ($|p_{e,z}| \simeq \mu_e$).

From our numerical data, we obtain the following order-of-magnitude estimate for the ratio of the net momentum to the energy emission rates:
\begin{eqnarray}
\eta \equiv \frac{ \dot{\cal P}_{\nu,z}}{\dot{\cal E}_{\nu}} \sim 2 \times 10^{-3} \frac{|eB|}{\mu_e T},
\label{def-eta}
\end{eqnarray} 
which quantifies the momentum asymmetry of the emitted neutrinos. For representative parameters, $\mu_e = 50~\mbox{MeV}$ and $T = 1~\mbox{MeV}$, this asymmetry reaches only the percent level for magnetic field strengths as large as $B \simeq 4 \times 10^{16}~\mbox{G}$. Moreover, even in the extreme-field limit where all occupied electron states reside in the LLL, the asymmetry remains moderate, with a peak value $\eta \lesssim 0.15$. These values are significantly smaller than those predicted by simplified models that estimate neutrino emission asymmetry based on electron spin polarization alone~\cite{Sagert:2007as,Ayala:2018kie,Ayala:2024wgb}.

\subsection{Estimate of pulsar kick velocity}
\label{sec:kick-velocity}

Since the background magnetic field induces a nonzero net momentum emission, neutrinos may, in principle, contribute to pulsar kicks. An estimate of the corresponding kick velocity, $v_k$, is given by
\begin{eqnarray}
 v_k=\frac{4\pi R_c^3}{3 M}\int \dot{\cal P}_{\nu,z}dt
 =\frac{4\pi R_c^3}{3 M} \int \eta \,  \dot{\cal E}_{\nu}dt ,
 \label{v-kick-def}
 \end{eqnarray} 
 where $R_c$ denotes the radius of the quark core and $M$ is the neutron-star mass. In the last expression, the momentum emission rate has been expressed in terms of the corresponding energy emission rate using the function $\eta$ introduced in Eq.~(\ref{def-eta}).

Assuming that the neutrino emission is the main cooling mechanism, the corresponding rate $\dot{\cal E}_\nu$ determines the cooling rate of the stellar quark core, 
\begin{eqnarray}
\frac{dT}{dt}=-\frac{\dot{\cal E}_\nu}{C_V}.
\label{cooling-rate}
\end{eqnarray}
where $C_V$ is the specific heat of dense quark matter. At small temperatures, its analytical expression is given by~\cite{Glendenning:1997wn}
\begin{eqnarray}
C_V(T)= 3 (\mu_u^2+\mu_d^2) T\left(1-\frac{2\alpha_s}{\pi}\right) +O\left(T^3\right),
\end{eqnarray}
with subleading terms suppressed by a factor on the order of $T^2/\mu_f^2$. Note that the electron contribution to the specific heat can be neglected here, as it is suppressed by a factor of the order of $(\mu_e/\mu_u)^2$. 

Using Eq.~(\ref{cooling-rate}) to change the integration variable from time $t$ to temperature $T$ in Eq.~(\ref{v-kick-def}), we obtain 
\begin{eqnarray}
 v_k  = - \frac{4\pi R_c^3}{3 M} \int_{T_i}^{T_f} \eta \, C_V(T')dT^\prime \simeq 8\pi  \times 10^{-3} \frac{R_c^3}{M_{\odot}} \frac{|eB|}{\mu_e} (\mu_u^2+\mu_d^2) \left(1-\frac{2\alpha_s}{\pi}\right)(T_i-T_f),
\label{v-kick-eq}
\end{eqnarray} 
where $T_i$ and $T_f$ denote the initial and final temperatures of the cooling stage. For typical compact stars with quark-core masses of the order of the solar mass, $M \simeq M_{\odot}$, core radii $R_c\simeq 10~\mbox{km}$, the resulting numerical estimate yields a pulsar kick velocity of only a few kilometers per second, even for temperature differences as large as $T_i-T_f \simeq 10~\mbox{MeV}$. This is clearly insufficient to account for the observed pulsar kick velocities, which are typically in the range $100$ to $1000~\mbox{km/s}$.

\section{Neutrino-antineutrino synchrotron emission}
\label{sec:synchrotron}

Unlike the direct Urca process, neutrino-antineutrino synchrotron emission, $q_f \rightarrow q_f + \nu_i + \bar{\nu}_i$, occurs in dense quark matter only in the presence of a magnetic field. The background field accelerates charged fermions, such as quarks or electrons, allowing them to radiate $\nu\bar{\nu}$ pairs. This mechanism is analogous to ordinary synchrotron photon emission, but instead proceeds via neutral $Z$-boson exchange and results in the emission of a neutrino-antineutrino pair, as illustrated by the Feynman diagram in Fig.~\ref{fig.Feynman}(b).

Here we review the derivation of this process using the Kadanoff–Baym formalism, as developed in Ref.~\cite{Ghosh:2025vkm} and briefly summarized in Sec.~\ref{sec:Kadanoff-Baym-eq}. Rather than computing particle amplitudes directly from wave functions \cite{Kaminker:1992su}, the derivation is formulated in terms of the neutrino self-energy. Working entirely with Green's functions, this approach provides a conceptually transparent and computationally efficient framework for evaluating neutrino emissivity in dense, magnetized quark matter.

Both quarks and electrons can contribute independently to the total neutrino luminosity from synchrotron emission. For clarity, we first consider a single quark flavor emitting a single neutrino-antineutrino pair via the weak interaction shown in Fig.~\ref{fig.Feynman}(b). This simplified setup captures the essential physics while allowing for compact analytical expressions. The formalism can then be readily generalized to astrophysical applications, such as quark stars, by including all quark flavors, all neutrino species, and the electron sector. As demonstrated below, the electron contribution reproduces the well-known results obtained using the standard amplitude-based formalism \cite{Kaminker:1992su}.

\subsection{Neutrino number production rate}
\label{sec:synch-number}

\begin{figure}[tbh]
\centering
\includegraphics[width=0.375\textwidth]{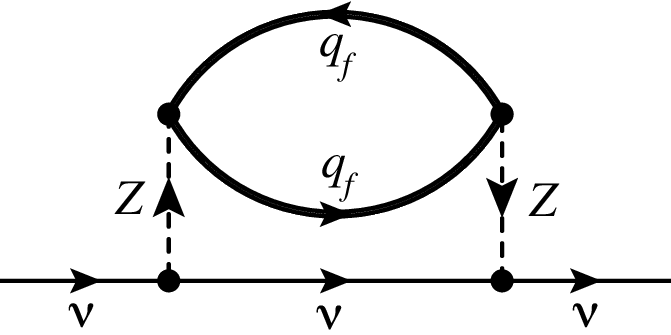}
\caption{Neutrino self-energy diagram used in the Kadanoff-Baym formalism to compute the neutrino-antineutrino synchrotron emission rate in dense quark matter.}
\label{fig:SelfEnergyZ}
\end{figure}

Using the interaction Lagrangian density in Eq.~(\ref{interaction-synchrotron}), we obtain the following contribution to the neutrino self-energy mediated by neutral $Z$-boson exchange 
\begin{equation}
 \Sigma^{\lessgtr}_\nu(P) =   i  \frac{G_F^2}{2} \int \frac{d^4 Q}{(2\pi)^4} \gamma^\delta (1-\gamma^5)
 G_\nu^{\lessgtr}(Q)\gamma^\sigma(1-\gamma^5){\Pi}^{\lessgtr}_{\delta\sigma}(P-Q) ,
 \label{Sigma-gtr}
\end{equation}
where ${\Pi}^{\lessgtr}_{\delta\sigma}(P-Q) $ are the lesser and greater $Z$-boson self-energies. The corresponding diagrammatic representation of this contribution is shown in Fig.~\ref{fig:SelfEnergyZ}.

We emphasize that, although Eqs.~(\ref{Sigma-gtr-1}) and (\ref{Sigma-gtr}) share a similar structure, they arise from distinct physical processes: the former corresponds to charged-current direct Urca emission mediated by $W$-boson exchange, while the latter describes neutrino synchrotron emission via neutral-current $Z$-boson exchange.

By substituting the Green's functions given in Eqs.~(\ref{G-less}) and (\ref{G-greater}) into Eq.~(\ref{Sigma-gtr}) and neglecting neutrino trapping (i.e., assuming that the neutrino and antineutrino distribution functions vanish), we obtain the following expression for the self-energies:
\begin{equation}
\Sigma^{\lessgtr}_\nu(P) =   \frac{G_F^2}{4} \int \frac{d^4 Q}{(2\pi)^3 q_{0}} \gamma^\delta (1-\gamma^5) \gamma^\lambda Q_{\lambda} \frac{1-\gamma_5}{2}\delta(q_{0}\pm q)\gamma^\sigma(1-\gamma^5){\Pi}^{\lessgtr}_{\delta\sigma}(P-Q).
\label{Sigma_exp}
\end{equation}
Finally, by inserting this expression into the Kadanoff-Baym transport  equation (\ref{KB-kinetic-eq}), we derive the neutrino-number production rate 
\begin{equation}
\frac{\partial f_\nu(t,\bm{p})}{\partial t} = \frac{G_F^2}{4} \int \frac{d^3 \bm{q}}{(2\pi)^3 p_{0}q_{0}}  L^{\delta\sigma}(Q,P)
n_B(p_{0}+q_{0}) \mbox{Im} \left[\Pi^R_{\delta\sigma}(P+Q)\right] ,
\label{d-f_nu-dt}
\end{equation}
where $P_{\lambda} = (p_{0},\bm{p})$ and $Q_{\lambda} = (q_{0},\bm{q})$ denote the four-momenta of the neutrino and antineutrino, respectively, with the on-shell conditions $p_{0}=p$ and $q_{0}=q$ imposed. In deriving Eq.~(\ref{d-f_nu-dt}) from Eq.~(\ref{Sigma_exp}), the lesser and greater $Z$-boson self-energies were expressed in terms of the imaginary part of the retarded self-energy, as given in Eqs.~(\ref{Pi-gtr}) and~(\ref{Pi-less}). Also, the change of variables $\bm{q} \to - \bm{q}$ was performed. The shorthand notation $L^{\delta\sigma}(Q,P)$ in Eq.~(\ref{d-f_nu-dt}) stands for the lepton tensor, which is defined by the following Dirac trace:
\begin{eqnarray}
L^{\delta\sigma}(Q,P)&=&
\mbox{Tr}\left[(p_0\gamma^0-\bm{\gamma}\cdot \bm{p})  \gamma^\delta (1-\gamma^5) (q_0\gamma^0-\bm{\gamma}\cdot \bm{q})  \gamma^\sigma(1-\gamma^5)\right] \nonumber\\
&=& 8\left[ Q^\delta P^\sigma+P^\delta Q^\sigma-g^{\delta\sigma}(P^{\rho}Q_{\rho})
+i\epsilon^{\delta\sigma \kappa\rho}Q_\kappa P_\rho
\right] ,
\end{eqnarray}
where $\epsilon^{\delta\sigma \kappa\rho}$ denotes the conventional Levi-Civita tensor.

Note that an expression analogous to Eq.~(\ref{d-f_nu-dt}) can be also derived for the antineutrino production rate $\partial_t f_{\bar{\nu}}(t,\bm{p})$. As expected, it produces an identical result.

\subsection{Neutrino energy emission rate}
\label{sec:synch-energy}

In terms of the neutrino and antineutrino distribution functions, the synchrotron emission rate is defined as
\begin{equation}
\dot{\cal E}_\nu = \int \frac{d^3\bm{p} }{(2\pi)^3}  p_0 \frac{\partial f_\nu(t,\bm{p} )}{\partial t}
+ \int \frac{d^3\bm{q} }{(2\pi)^3}  q_0 \frac{\partial f_{\bar{\nu}}(t,\bm{q} )}{\partial t}.
\label{def-rate}
\end{equation}
By making use of the neutrino-number production rate in Eq.~(\ref{d-f_nu-dt}) and noting that the antineutrino rate is the same, we then derive
\begin{equation}
\dot{\cal E}_\nu =  \frac{G_F^2}{4} \int \frac{d^3\bm{p} d^3 \bm{q}}{(2\pi)^6 q_0 p_0} 
 (p_0 +q_0) n_B( p_0+q_0) L^{\delta\sigma}(Q,P) \mbox{Im} \left[\Pi^R_{\delta\sigma}(P+Q)\right].
\label{E-rate}
\end{equation}
This latest form can be  further simplified by making use of the following identity \cite{Ghosh:2025vkm}:
\begin{eqnarray}
 \int  \frac{d^3\bm{p} d^3 \bm{q} }{q p} f(Q+P) Q^{\mu} P^{\nu}=\frac{\pi}{6}\int_{0}^{\infty} d\tilde{q}_0 \int d^3 \tilde{\bm{q}} \, 
 \theta(\tilde{q}_0^2-\tilde{q}^2) f(\tilde{Q})  \left( g^{\mu\nu} \tilde{Q}^2+2 \tilde{Q}^{\mu} \tilde{Q}^{\nu} \right) ,
 \label{integral-trick}
\end{eqnarray}
where $f(Q)$ is an arbitrary function. Here we used the following notation for the four-vectors: $Q=(q,\bm{q})$, $P=(p,\bm{p})$, and $\tilde{Q}\equiv (\tilde{q}_0,\tilde{\bm{q}})=Q+P$. The four-vector $\tilde{Q}$ represents the total energy ($\tilde{q}_0=p_0+q_0$) and momentum ($\tilde{\bm{q}}=\bm{p}+\bm{q}$) of the neutrino-antineutrino pair produced in a single weak process, $q_f \to q_f + \nu + \bar{\nu}$. Since neutrinos are assumed massless here, the four-momentum $\tilde{Q}$ is timelike: $\tilde{Q}^2=2q_0 p_0 (1-\cos\varphi_{\nu\bar{\nu}})\geq 0$, where $\cos\varphi_{\nu\bar{\nu}}$ denotes the angle between the $\nu$ and $\bar{\nu}$ momenta.

We henceforth drop the tilde in $\tilde{Q}$ and denote the total four-momentum of the neutrino-antineutrino pair as $Q$. Then, the neutrino energy emission rate reads
\begin{equation}
\dot{\cal E}_\nu = -  \frac{2 G_F^2}{3(2\pi)^5}
\int_{0}^{\infty} dq_0 \int_{q^2\leq q_0^2} d^3\bm{q} \, q_0 n_B( q_0) 
 \left( g^{\delta\sigma} Q^2 -  Q^{\delta} Q^{\sigma} \right) 
\mbox{Im} \left[\Pi^R_{\delta\sigma}(Q)\right],
\label{E-dot-1}
\end{equation}
where the timelike nature of $Q$ is enforced by limiting the integration range to $q^2\leq q_0^2$.

Using the Landau-level representation for the quark propagators in a background magnetic field, it is straightforward to derive the imaginary part of the retarded self-energy, $\mbox{Im} \left[\Pi^R_{\delta\sigma}(Q)\right]$, and contract it with the the lepton tensor. For details of the derivation, see  Appendix~A of Ref.~\cite{Ghosh:2025vkm}. Using the corresponding result, we obtain the following expression for the rate:
\begin{eqnarray}
\dot{\cal E}_\nu &=& \frac{N_c G_F^2}{6 (2\pi)^6 \ell^2}  \sum_{n=0}^{\infty}  \sum_{s=1}^{\infty} 
\int_{-\infty}^{\infty} d k_z \int_{q^2\leq q_0^2}  d^3\bm{q} \,q_0
\frac{ n_F(E_{n^\prime,k_z+q_z}-\mu_f) \left[1-n_F(E_{n,k_z}-\mu_f) \right] }{E_{n^\prime,k_z+q_z} E_{n,k_z}} 
\nonumber \\
&\times& 
 \left[\left((c_V^f)^2+(c_A^f)^2\right) \tilde{S}_1 +\left((c_V^f)^2-(c_A^f)^2\right) \tilde{S}_2 + c_V^f c_A^f \tilde{S}_3\right] ,
\label{E-dot-30}
\end{eqnarray}
where $n_F(E-\mu)\equiv 1/\left[e^{(E-\mu)/T}+1\right]$ denotes the Fermi-Dirac distribution function, $E_{n,k_z} = \sqrt{2n |e_f B|+k_z^2+m^2}$ is the quark Landau-level energy, $\ell = 1/\sqrt{|e_f B|}$ is the magnetic field length, and $s=n^\prime-n\geq 1$ is the difference between the Landau-level indices of the initial and final quark states. The total energy of the neutrino and antineutrino pair is given by $q_0 \equiv E_{n^\prime,k_z+q_z} - E_{n,k_z}$. 

The explicit expressions for the functions $\tilde{S}_i$ are given by:
\begin{eqnarray}
\tilde{S}_1&=& \left[-2 (q_0^2-q_z^2-q_\perp^2) \frac{n+n^\prime}{\ell^2} 
-m^2\left(q_0^2-q_z^2\right) \right]
\left(\mathcal{I}_{0}^{n,n^\prime}(\xi_q) +\mathcal{I}_{0}^{n-1,n^\prime-1}(\xi_q)\right) \nonumber\\
&+&\left[2 (q_0^2-q_z^2-q_\perp^2) \frac{n+n^\prime}{\ell^2} 
- \left(q_0^2-q_z^2 -q_\perp^2 \right)^2 
+m^2 \left(2q_0^2-2q_z^2-q_\perp^2\right)
\right] \nonumber\\
&\times& \left(\mathcal{I}_{0}^{n,n^\prime-1}(\xi_q) +\mathcal{I}_{0}^{n-1,n^\prime}(\xi_q)\right) , 
\label{S1-best}  \\
\tilde{S}_2&=&- m^2 \left(\mathcal{I}_{0}^{n,n^\prime}(\xi_q)+\mathcal{I}_{0}^{n-1,n^\prime-1}(\xi_q)\right) \left(q_0^2-q_z^2-2 q_\perp^2\right) \nonumber\\
&-& m^2 \left( \mathcal{I}_{0}^{n,n^\prime-1}(\xi_q)+\mathcal{I}_{0}^{n-1,n^\prime}(\xi_q)\right)  \left(2q_0^2-2q_z^2-q_\perp^2  \right)  ,   
\label{S2-best}  \\
\tilde{S}_3&=& 2 s_\perp  \left[ (k_z+q_z)E_{n} -k_z E_{n^\prime} \right] \left[\frac{2 (n^\prime-n)}{\ell^2} \left(\mathcal{I}_{0}^{n,n^\prime}(\xi_q)-\mathcal{I}_{0}^{n-1,n^\prime-1}(\xi_q)\right) \right. \nonumber\\
&+& \left .\left( \mathcal{I}_{0}^{n,n^\prime-1}(\xi_q)-\mathcal{I}_{0}^{n-1,n^\prime}(\xi_q)\right)  \left(2q_0^2-2q_z^2 - 3 q_\perp^2\right) 
 \right] ,  
 \label{S3-best}
\end{eqnarray}
where $\xi_q =\bm{q}_\perp^2\ell^2/2$, and the form-factor function $\mathcal{I}_{0}^{n,n^\prime}\left(\xi\right)$ is defined in terms of generalized Laguerre polynomials as follows \cite{Wang:2021ebh}:
\begin{equation}
\mathcal{I}_{0}^{n,n^{\prime}}(\xi) =  \frac{n!}{(n^\prime)!}e^{-\xi} \xi^{n^\prime-n} \left(L_{n}^{n^\prime-n}\left(\xi\right)\right)^2 .
\label{I0-nnprime}  
\end{equation}
The expression in Eq.~(\ref{E-dot-30}) constitutes the final result for the $\nu\bar{\nu}$ synchrotron emission rate. As can be readily verified, it is consistent with the corresponding result obtained using a different approach in Ref.~\cite{Kaminker:1992su}. We note, however, that our representation of the function $\tilde{S}_1$ is more compact, as it makes explicit use of the definition of $q_0$ together with the Landau-level energy spectrum.

Although the synchrotron emission rate in Eq.~(\ref{E-dot-30}) is exact, it is not the most convenient for applications to dense quark matter under compact-star conditions. This is a consequence of a large hierarchy of energy scales, set by the quark chemical potential $\mu_f$ relative to the temperature $T$ and the Landau-level spacing $\delta\epsilon_B$, which results in a Landau-level sum containing a large number of terms. Physically, this reflects the contribution of many quantum transitions among states near the Fermi surface.

Using the two relevant low-energy scales, i.e., the temperature $T$ and  the Landau-level spacing at the Fermi surface, $\delta\epsilon_B \equiv |e_f B|/\mu_f$, it is natural to define two qualitatively distinct regimes: the weak-field, high-temperature regime ($b \ll 1$) and the strong-field, low-temperature regime ($b \gg 1$).

In the weak-field regime, many transitions between closely spaced Landau levels ($s_{\rm max} \gg 1/b$) contribute, effectively suppressing Landau-level quantization effects. However, the rate vanishes as $B \to 0$ due to kinematic constraints. In contrast, in the strong-field regime, Landau-level quantization dominates, and the rate is controlled by transitions between well-separated adjacent levels ($s \gtrsim 1$), leading to an exponential suppression as the density of thermally excited quarks near the Fermi surface decreases.

Since the Landau level separation at the Fermi surface $\delta\epsilon_B \equiv |e_f B|/\mu_{f} $ is much smaller than the quark chemical potential itself, the sum over index $n$ can be approximated by an integral, i.e.,
\begin{eqnarray}
\sum_{n=0}^{\infty}  f(2n|e_f B|) = \frac{\ell^2}{2\pi} \int d^2 \bm{k}_\perp f(k_\perp^2)  .
\end{eqnarray}
Then the emission rate takes the following form:  
\begin{eqnarray}
\dot{\cal E}_\nu &=& \frac{N_c G_F^2}{6 (2\pi)^6} \sum_{s=1}^{\infty} 
\int_0^{\infty}  k^2 dk \int_{0}^{\pi}  \sin\theta d\theta \int_{q^2\leq q_0^2}  d^3\bm{q} \,q_0
\frac{ n_F(E_{k}+q_0-\mu_{f} ) \left[1-n_F(E_{k}-\mu_{f} ) \right] }{E_{k} (E_{k}+q_0)} 
\nonumber \\
&\times& 
 \left[\left((c_V^f)^2+(c_A^f)^2\right) \tilde{S}_1 +\left((c_V^f)^2-(c_A^f)^2\right) \tilde{S}_2 + c_V^f c_A^f \tilde{S}_3\right] ,
\label{E-dot-40}
\end{eqnarray}
where $q_0 \simeq \left(k q_z \cos\theta+s |e_f B|\right)/E_k$, $E_{k}=\sqrt{k^2+m^2}$, $k_z =k \cos\theta$, $k_\perp =k \sin\theta$, and the argument of the Bessel functions $2\sqrt{\xi_q n}$ is replaced with $k_\perp q_\perp \ell^2$. In the last equation, functions $\tilde{S}_i$ are given by the same expressions as before, see Eqs.~(\ref{S1-best})--(\ref{S3-best}), except that the form factor functions $\mathcal{I}_{0}^{n,n+s}(\xi_q)$ are replaced by the following large-$n$ asymptotes \cite{Szego:1975}:
\begin{equation}
\mathcal{I}_{0}^{n,n+s}(\xi) \simeq \left[J_{s}\left(2\sqrt{\xi n}\right)\right]^2, 
\label{I00-approx}
\end{equation}
where $s\ll n$ is assumed. 

Since the rate is dominated by quark states in the vicinity of the Fermi surface, it is a good approximation to replace the quark momentum by its Fermi-surface value ($k \approx k_F \approx \mu_{f}$) throughout the integrand, except in the distribution functions. The resulting integrals over $k$ and $q_\perp$ can then be carried out analytically, yielding
\begin{eqnarray}
\dot{\cal E}_\nu&=&\frac{2 N_c G_F^2 |e_fB|^2 }{3  (2\pi)^5} \left((c_V^f)^2+(c_A^f)^2\right)  \sum_{s=1}^{\infty} 
\int_{0}^{\pi} \sin\theta d\theta \int_{-s \delta\epsilon_B/(1+\cos\theta)}^{s \delta\epsilon_B/(1-\cos\theta)} dq_z  
 \nonumber \\
 &\times& q_0^2 \,  n_B(q_0)(q_0^2-q_z^2) J_{s+1}\left(w\right)J_{s-1}\left(w\right),
\label{E-dot-41a}
\end{eqnarray}
where $q_0 \simeq q_z \cos\theta +s \delta\epsilon_B$ and $w\equiv  \mu_{f}  \ell^2 \sqrt{q_0^2-q_z^2} \sin\theta$. This approximation captures all leading-order contributions while neglecting subleading terms suppressed by inverse powers of the quark chemical potential.

It is worth emphasizing that the rate in Eq.~(\ref{E-dot-41a}) does not scale with the Fermi-surface area, which is $\propto \mu_{f}^2$ \cite{Kaminker:1992su,Bezchastnov:1997ew}. Although one might naively expect such scaling, since the density of quark states participating in the $\nu\bar{\nu}$ synchrotron emission is proportional to the Fermi-surface area, this dependence is exactly compensated by the emission matrix element, which is inversely proportional to the squared momentum of the initial quark state, $1/k_F^2 \simeq 1/\mu_{f}^2$. As a result, the only remaining dependence on $\mu_{f} $ enters through the low-energy parameter $\delta\epsilon_B \equiv |e_f B|/\mu_{f} $. 

The final expression is convenient to render in the following form:
\begin{equation}
\dot{\cal E}_\nu = \frac{2 N_c G_F^2|e_fB|^2 T^5}{3  (2\pi)^5} \left((c_V^f)^2+(c_A^f)^2\right) F(b),
\label{E-dot-final}
 \end{equation}
where $b \equiv \delta\epsilon_B/T = |e_fB|/(T\mu_{f} )$ is the dimensionless ratio of the two low-energy energy scales introduced earlier.  The scaling function $F\left(b\right) $ is defined by
\begin{equation}
F\left(b\right) = \sum_{s=1}^{\infty} 
 \int_{-1}^{1}dy \int_{-sb/(1+y)}^{s b/(1-y)} d u_z \,
 \frac{\left(u_z y +s b\right)^2 \left[ \left(u_z y +s b\right)^2-u_z^2\right] }{e^{u_z y +s b}-1} J_{s+1}\left(\tilde{w}\right)J_{s-1}\left(\tilde{w}\right) ,
\label{F-b}
 \end{equation}
with $\tilde{w}=\frac{1}{b} \sqrt{1-y^2} \sqrt{\left(y u_z +s b\right)^2-u_z^2}$. Here we changed the integration variable from $\theta$ to $y=\cos\theta$, and introducing the dimensionless variables $u_z \equiv q_z/T$ and $u_0 \equiv q_0/T = u_z  y +s b$. 

\subsection{Numerical results for $\nu\bar{\nu}$ synchrotron emission rate}
\label{sec:synch-numerical}

Here we use the final expression in Eq.~(\ref{E-dot-final}) to compute numerically the $\nu\bar{\nu}$ synchrotron emission rate in dense quark matter. By evaluating the function $F(b)$ numerically, we find that it decreases monotonically with $b$ and can be accurately approximated by the fit
\begin{equation}
\label{F-fit}
F(b)\approx
\frac{11.06 - 0.46 b + 1.13 b^2 + 0.017 b^3 + 0.00165 b^4}
{1 - 0.022 b + 1.4\times10^{-4} b^3 + 7.7\times10^{-5} b^4 + 1.5\times10^{-7} b^5}
\,e^{-b/2}.
\end{equation}
valid over the range $0\lesssim b\lesssim100$ (with an accuracy of approximately $\sim3\%$). Numerical values of the function $F\left(b\right) $ are provided in Ref.~\cite{DataF:2025}. 

\begin{figure}[t]
\centering
\includegraphics[width=0.9\textwidth]{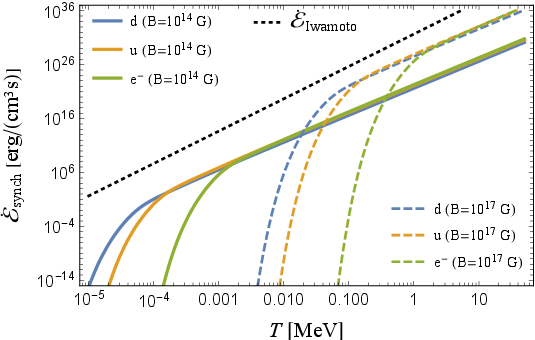}
\caption{Temperature dependence of the synchrotron neutrino emission rates from quarks and electrons in two-flavor dense quark matter for two magnetic field strengths: $B=10^{14}~\mbox{G}$ (solid lines) and $B=10^{17}~\mbox{G}$ (dashed lines). For comparison, Iwamoto's direct Urca emission rate is also plotted (black dotted line).\label{fig:rates-vs-T}}
\end{figure}

The temperature dependence of synchrotron emission for each type of charged fermion is presented in Fig.~\ref{fig:rates-vs-T}. To highlight the role of the magnetic field, numerical results are shown for two field strengths: $B=10^{14}~\mbox{G}$ (solid lines) and $B=10^{17}~\mbox{G}$ (dashed lines). For comparison, we also plot Iwamoto's direct Urca emission rate (black dotted line) \cite{Iwamoto:1980eb,Iwamoto:1982zz}, defined by Eq.~(\ref{dot-E-app-Iwamoto}). 

As illustrated in the figure, synchrotron emission rates from both quarks and electrons are suppressed by several orders of magnitude relative to the direct Urca rate, even at an extreme magnetic field strength of $B = 10^{17}~\mbox{G}$. It is quite interesting to note also that the electron synchrotron contribution becomes comparable to that of quarks, despite the electron number density being extremely small.

At first sight, the $T^5$ temperature scaling of the synchrotron emission rate in Eq.~(\ref{E-dot-final}) suggests that it could dominate over the direct Urca rate in Eq.~(\ref{dot-E-app-Iwamoto}), which scales as $T^6$, particularly at sufficiently low temperatures. This expectation is partially realized at high temperatures, as seen in Fig.~\ref{fig:rates-vs-T}, where the synchrotron rate exhibits a smaller slope. However, this apparent advantage is lost once the temperature drops below the Landau-level spacing scale, $T\lesssim |e_fB|/\mu_{f}$, at which point the synchrotron emission rate becomes exponentially suppressed.

\section{Summary and Outlook}
\label{sec:discussion}

We have investigated neutrino production in strongly magnetized, unpaired dense quark matter, focusing on number, energy, and momentum emission rates, using first-principles field-theory methods. Employing the Kadanoff-Baym formalism, we derived the neutrino number production rate and obtained from it the corresponding energy and momentum emission rates. Motivated by compact-star applications, we considered the regime in which quark chemical potentials greatly exceed that of electrons, retaining Landau quantization only for electrons while treating quarks in the high-density (many-Landau-level) limit.

The Landau-level discretization of electron states near the Fermi surface induces an oscillatory dependence of the neutrino emission rates on the magnetic-field strength for $|eB| \gtrsim \pi T \mu_e$, with the peaks at $|eB| = \mu_e^2/(2n)$. The oscillation amplitude increases as the temperature decreases and formally diverges in the limit $T \to 0$. However, thermal broadening suppresses these effects at weaker fields or higher temperatures. For $B \lesssim 10^{17}~\mathrm{G}$, the oscillations are negligible at $T \gtrsim 2~\mathrm{MeV}$. On average, the energy emission rate decreases with increasing field strength, although the suppression remains modest, at the level of $\lesssim 20\%$. The average net momentum emission along the field direction increases with $B$ but remains small. In particular, the resulting anisotropy, quantified by $\eta \equiv \dot{\mathcal{P}}_{\nu,z}/\dot{\mathcal{E}}_{\nu}$, reaches only the percent level for $B \lesssim 10^{16}~\mathrm{G}$, corresponding to kick velocities of at most a few kilometers per second, which is far below those required to account for observed pulsar kicks.

We have also reviewed $\nu\bar{\nu}$ synchrotron emission in magnetized two-flavor quark matter. The analytic expression for the emissivity is given in Eq.~(\ref{E-dot-final}) and, apart from its overall scaling with magnetic field strength and temperature, $\propto |eB|^{2}T^{5}$, it is fully determined by a universal function $F(b)$ of the single dimensionless parameter $b=|e_fB|/(T\mu_f)$. In the weak-field limit, $b\ll1$, the rate receives contributions from transitions among many Landau levels. In contrast, in the strong-field limit, $b\gg1$, the emission is dominated by transitions between adjacent levels but is exponentially suppressed. A convenient numerical fit for the function $F(b)$ is provided in Eq.~(\ref{F-fit}). Including the electron contribution, the total synchrotron emissivity remains several orders of magnitude smaller than the direct Urca emissivity for magnetic fields up to $10^{17}~\mathrm{G}$ and for typical stellar temperatures. Thus, it is unlikely to compete with direct Urca processes as a cooling mechanism in realistic quark cores of compact stars.

As an outlook for future work, at least two directions warrant further investigation. First, the early deleptonization stage of compact stars, in which neutrino trapping and diffusion are essential, has not been addressed in the present study. In this regime, anisotropic neutrino diffusion in a strongly magnetized, lepton-rich medium may generate momentum asymmetries qualitatively different from those found in the free-streaming limit considered here. This requires detailed microscopic calculations at nonzero neutrino chemical potential and self-consistent transport modeling.

Second, since quark matter is expected to form Cooper pairs and enter a color-superconducting state at sufficiently high densities~\cite{Shovkovy:2004me,Alford:2007xm}, it is important to investigate the impact of diquark pairing on neutrino emission and the associated momentum asymmetries. Without detailed calculations, one can argue that the results should depend sensitively on the underlying diquark pairing pattern~\cite{Schafer:2004jp}.

In the two-flavor color-superconducting (2SC) phase, a third of quark modes remain ungapped, so its emission should resemble that of normal quark matter, with the rate about three times smaller \cite{Jaikumar:2005hy,Alford:2025tbp}. Indeed the contribution of gapped modes is exponentially suppressed at low temperatures, scaling as $\dot{\cal E}_\nu \propto e^{-\Delta/T}$, where $\Delta \sim \mathcal{O}(10\,\mathrm{MeV})$ denotes the color-superconducting gap. At sufficiently low temperatures, such modes therefore contribute very little.

By contrast, in the color-flavor-locked (CFL) phase all quark modes are gapped, strongly suppressing low-energy excitations and rendering negligible neutrino emission at low temperatures. Notably, the specific heat of CFL phase is also exponentially suppressed and, thus, the cooling of compact stars with quark cores in the CFL phase is a delicate issue \cite{Shovkovy:2002kv}.

Finally, single-flavor spin-one color-superconducting phases admit multiple possible pairing patterns, each requiring separate analysis. Of particular interest are those phases in which the gap exhibits nodes or lines on the Fermi surface. Such gapless phases permit low-energy excitations and can therefore lead to non-negligible neutrino emission with potentially nontrivial consequences for stellar cooling \cite{Schmitt:2005wg}.

These general arguments established for color superconductors in the zero-field limit may remain qualitatively valid in the presence of weak magnetic fields. However, the interplay between color superconductivity and strong magnetic fields is far from fully understood. Intense magnetic fields can alter the gap structure \cite{Ferrer:2005vd,Ferrer:2006vw,Noronha:2007wg,Fayazbakhsh:2010gc}, induce anisotropies in the pairing pattern \cite{Yu:2012jn}, and preferentially stabilize specific superconducting phases \cite{Feng:2009vt}. A systematic treatment of neutrino emission in magnetized color-superconducting quark matter is therefore crucial for developing reliable models of the thermal evolution and observable signatures of strongly magnetized compact stars.

\vspace{6pt} 


\authorcontributions{Conceptualization, I.S. and R.G.; methodology, I.S. and R.G.; formal analysis, I.S. and R.G.; investigation, I.S. and R.G.; writing -- original draft preparation, I.S. and R.G.; writing -- review and editing, I.S. and R.G.; funding acquisition, I.S. Both authors have read and agreed to the published version of the manuscript.}

\funding{This research was funded by the U.S. National Science Foundation under Grant Nos.~PHY-2209470 and PHY-2514933. R.G. was partly supported by Academia Sinica through Project No. AS-CDA-114-M01 and additionally by an Academia Sinica postdoctoral fellowship.}

\dataavailability{The numerical data for the function $F(b)$ determining the synchrotron emission rate are available as the Supplemental Material in Ref.~\cite{DataF:2025}.} 


\conflictsofinterest{The authors declare no conflicts of interest.} 



\abbreviations{Abbreviations}{
The following abbreviations are used in this manuscript:
\\

\noindent 
\begin{tabular}{@{}ll}
QED & Quantum electrodynamics\\
QCD & Quantum chromodynamics\\
LLL & Lowest Landau level\\
CFL & color-flavor-locked\\
2SC & two-flavor color-superconducting
\end{tabular}
}

\begin{adjustwidth}{-0.1\extralength}{0cm}

\reftitle{References}

\PublishersNote{}
\end{adjustwidth}
\end{document}